\documentclass[11pt]{article}

\usepackage{amsmath,amssymb}
\usepackage{adjustbox}
\usepackage{booktabs}
\usepackage[margin=1in]{geometry}
\usepackage{graphicx}
\usepackage[table]{xcolor}
\usepackage{tikz}
\usepackage{array}
\usepackage{multirow}
\usepackage{url}
\usepackage{cite}
\usepackage{caption}
\usepackage{setspace}
\usepackage[section]{placeins}
\usepackage{xspace}
\usetikzlibrary{arrows.meta,positioning,fit,calc}

\newcommand{\dataset}{LAMBDA\xspace}
\newcommand{\cmark}{\checkmark}
\newcommand{\xmark}{\(\times\)}
\newcommand{\pmark}{\(\triangle\)}
\newcommand{\parahead}[1]{\par\medskip\noindent\textbf{#1.}\ }
\newcommand{\taskhead}[1]{\par\medskip\noindent\textit{#1.}\ }
\begin{document}

\title{\dataset: A Low-Altitude Multimodal Base Dataset for UAV Sensing and Communication}

\author{
\parbox{0.95\textwidth}{\centering
Lin Zhou\textsuperscript{1},
Peichuan Rao\textsuperscript{1},
Chenshuo Zhang\textsuperscript{1},
Jianhua Mo\textsuperscript{1,*},
Shu Sun\textsuperscript{1},\\
Zhiyong Chen\textsuperscript{1},
and Meixia Tao\textsuperscript{1,*}}}
\date{}

\maketitle

\begingroup
\renewcommand{\thefootnote}{\arabic{footnote}}
\footnotetext[1]{School of Information Science and Electronic Engineering, Shanghai Jiao Tong University, Shanghai, China.}
\endgroup
\begingroup
\renewcommand{\thefootnote}{\fnsymbol{footnote}}
\footnotetext[1]{e-mail: mjh@sjtu.edu.cn; mxtao@sjtu.edu.cn}
\endgroup

\begin{abstract}
Research on low-altitude integrated sensing and communication (ISAC) requires aligned multimodal data that jointly describe wireless propagation, visual appearance, unmanned aerial vehicle (UAV) motion, light detection and ranging (LiDAR) perception, and radar sensing under common trajectories and timestamps.
To address this need, a \textbf{l}ow-\textbf{a}ltitude \textbf{m}ultimodal \textbf{b}ase \textbf{da}taset, named LAMBDA, is introduced.
LAMBDA is characterized by high fidelity, modality diversity, scenario richness, and configuration flexibility.
It is generated through a high-fidelity digital-twin pipeline with detailed scene geometry, refined material assignment, and electromagnetic modeling of UAVs.
LAMBDA provides synchronized RGB images, depth maps, LiDAR point clouds, inertial measurement unit states, UAV poses, channel state information (CSI), and radar-synthesis resources across matched low-altitude operating conditions, shared coordinate systems, and synchronized frame indices. The dataset covers urban, suburban, and campus scenes, multi-UAV/multi-base-station settings, nighttime conditions, and sunny, rainy, snowy, and foggy weather variations. Its CSI and radar resources support user-defined antenna-array sizes, bandwidths, subcarrier spacings, chirp parameters, and plane-wave or spherical-wavefront channel synthesis. The reliability and usability of LAMBDA are assessed through quality control, weather and multimodal visualization, and two UAV ISAC-related use cases: RGB-aided beam prediction and RGB--LiDAR-based UAV localization.
\end{abstract}

\section{Background \& Summary}
\label{sec:background-summary}
The rapid proliferation of unmanned aerial vehicles (UAVs) has enabled a wide range of low-altitude applications, including autonomous delivery, emergency rescue, and infrastructure inspection~\cite{mozaffari2019uavtutorial,shum2026lowaltitudeeconomy}. 
Unlike conventional terrestrial systems, low-altitude UAV networks involve small aerial targets, rapidly varying air--ground channels, three-dimensional mobility, and strong sensitivity to surrounding geometry and weather conditions. 
Safe, reliable, and efficient UAV operation therefore requires not only robust wireless communication links but also accurate environmental perception and motion awareness under highly dynamic and complex propagation and sensing conditions. 
These requirements naturally drive the convergence of communication and sensing functions on the same platforms and within the same networks, giving rise to a key research area known as integrated sensing and communication (ISAC)~\cite{liu2022isac,jiang2025isaclae,song2025cellularisac}.
ISAC is also recognized as a core capability of sixth-generation (6G) wireless networks~\cite{itu2023imt2030,alouini2024road6g}, enabling applications such as predictive beamforming, localization, UAV trajectory planning, and environment reconstruction. 
However, progress in these applications is still fundamentally constrained by the scarcity of comprehensive, high-quality multimodal datasets for low-altitude UAV scenarios. 
Such datasets should jointly capture wireless propagation, visual appearance, geometric structure, UAV motion, and weather-dependent variations under synchronized trajectories and common coordinate systems, thereby providing reliable foundations for algorithm validation, benchmarking, and model training.

\begin{table}[tb!]
\centering
\caption{Comparison of representative multimodal wireless and synthetic-city datasets. \cmark: included; \xmark: absent or not applicable; \pmark: limited or partially supported. }
\label{tab:dataset_comparison}
\footnotesize
\setlength{\tabcolsep}{2.4pt}
\renewcommand{\arraystretch}{1.08}
\resizebox{\textwidth}{!}{
\begin{tabular}{@{}lccccccccl@{}}
\toprule
Dataset & Year & \shortstack{Low-\\altitude} & \shortstack{CSI} & \shortstack{RGB/\\depth} & LiDAR & \shortstack{Radar} & \shortstack{IMU/\\GPS} & \shortstack{Weather/\\time} & Main focus \\
\midrule
KITTI~\cite{geiger2012kitti} & 2012 & \xmark & \xmark & \cmark & \cmark & \xmark & \cmark & \pmark & Terrestrial autonomous-driving perception \\
OPV2V~\cite{xu2021opv2v} & 2022 & \xmark & \xmark & \cmark & \cmark & \xmark & \pmark & \pmark & V2V cooperative perception \\
DAIR-V2X~\cite{yu2022dairv2x} & 2022 & \xmark & \xmark & \cmark & \cmark & \xmark & \pmark & \pmark & Real-world vehicle-infrastructure perception \\
SDCD~\cite{li2024sdcd} & 2024 & \xmark & \xmark & \cmark & \xmark & \xmark & \xmark & \cmark & Synthetic digital-city RGB-depth robustness \\
DeepMIMO~\cite{alkhateeb2019deepmimo} & 2019 & \xmark & \cmark & \xmark & \xmark & \xmark & \xmark & \pmark & Configurable ray-tracing channels \\
WAIR-D~\cite{huangfu2022waird} & 2022 & \xmark & \cmark & \xmark & \xmark & \xmark & \xmark & \xmark & Wireless AI channels over real-world maps \\
ViWi~\cite{alrabeiah2020viwi} & 2020 & \xmark & \cmark & \cmark & \pmark & \xmark & \xmark & \pmark & Vision-aided wireless communications \\
E-FLASH~\cite{gu2022multimodality} & 2022 & \xmark & \pmark & \pmark & \cmark & \xmark & \cmark & \pmark & Real-world mmWave V2X beam selection \\
DeepSense 6G~\cite{alsabah2023deepsense} & 2023 & \pmark & \pmark & \cmark & \cmark & \cmark & \pmark & \pmark & Real-world multimodal wireless measurements \\
DeepVerse 6G~\cite{wilabdeepverse} & 2024 & \pmark & \cmark & \cmark & \pmark & \cmark & \pmark & \pmark & Digital-twin wireless datasets \\
    M\textsuperscript{3}SC~\cite{cheng2023m3sc} & 2023 & \xmark & \cmark & \cmark & \cmark & \cmark & \xmark & \cmark & Mixed multimodal ISAC data \\
SynthSoM~\cite{cheng2025synthsom} & 2025 & \pmark & \cmark & \cmark & \cmark & \cmark & \xmark & \cmark & Synthetic SoM dataset with air-ground scenarios \\
Multimodal-Wireless~\cite{mao2026multimodalwireless} & 2025 & \xmark & \cmark & \cmark & \cmark & \cmark & \cmark & \cmark & V2X multimodal communication and perception \\
Multimodal-NF~\cite{li2026multimodalnf} & 2026 & \cmark & \cmark & \cmark & \cmark & \xmark & \cmark & \pmark & Near-field low-altitude XL-MIMO \\
PML-CellularEye~\cite{zhong2026pmlcellulareye} & 2026 & \cmark & \pmark & \pmark & \xmark & \pmark & \cmark & \cmark & Real-world BS-side low-altitude ISAC data \\
\dataset (this work) & 2026 & \cmark & \cmark & \cmark & \cmark & \cmark & \cmark & \cmark & Low-altitude multimodal ISAC data\\
\bottomrule
\end{tabular}
}
\end{table}

Datasets related to perception, wireless communication, and sensing have evolved along several related but distinct directions, yet existing representative datasets still do not fully meet the requirements of low-altitude UAV ISAC research.
Autonomous-driving and synthetic-city datasets, such as KITTI~\cite{geiger2012kitti}, OPV2V~\cite{xu2021opv2v}, DAIR-V2X~\cite{yu2022dairv2x}, and SDCD~\cite{li2024sdcd}, provide benchmarks with camera, light detection and ranging (LiDAR), GPS/inertial measurement unit (IMU), cooperative-perception, or weather-varied RGB-depth data.
However, these datasets are primarily designed for ground-centric agents, viewpoints, and motion patterns, and therefore cannot capture infrastructure-side observations of small aerial targets in low-altitude airspace.
Moreover, they generally lack radio-frequency (RF) channel labels, making them insufficient for ISAC studies. 
Wireless-channel datasets, such as DeepMIMO~\cite{alkhateeb2019deepmimo} and WAIR-D~\cite{huangfu2022waird}, provide ray-tracing-based RF channels over realistic maps and support communication-oriented evaluation. 
Nevertheless, their channel records are not synchronized with visual, geometric, LiDAR, inertial, and pose modalities, and their scenarios are mainly designed for terrestrial wireless links rather than low-altitude UAV sensing and communication. 

More recently, multimodal sensing-communication datasets have started to narrow this gap. 
ViWi~\cite{alrabeiah2020viwi} integrates visual scenes with wireless channels, while E-FLASH~\cite{gu2022multimodality} and DeepSense 6G~\cite{alsabah2023deepsense} provide real-world multimodal measurements for beam selection and other wireless tasks. 
M\textsuperscript{3}SC~\cite{cheng2023m3sc}, SynthSoM~\cite{cheng2025synthsom}, DeepVerse 6G~\cite{wilabdeepverse}, Multimodal-Wireless~\cite{mao2026multimodalwireless}, Multimodal-NF~\cite{li2026multimodalnf}, and PML-CellularEye~\cite{zhong2026pmlcellulareye} further advance multimodal sensing-communication data generation through simulation, hybrid generation, or real-world cellular testbeds. 
However, these datasets still provide only partial coverage of the low-altitude UAV ISAC design space. 
As summarized in Table~\ref{tab:dataset_comparison}, they do not jointly offer high-fidelity low-altitude digital-twin scenes, synchronized visual/geometric/inertial/wireless records, diverse UAV trajectories under controllable weather and time-of-day conditions, and configurable CSI and radar-generation support. 

To address this gap, \dataset, short for \textbf{L}ow-\textbf{A}ltitude \textbf{M}ultimodal \textbf{B}ase \textbf{DA}taset, is introduced as a foundational low-altitude data resource that can serve as a benchmark for UAV communication, sensing, tracking, and multimodal wireless model pretraining. 
\dataset{} is generated on a digital-twin platform that integrates Unreal Engine 5 (UE5)~\cite{unrealengine5} for high-fidelity scene construction; Cosys-AirSim~\cite{shah2017airsim,cosysairsim2023} for UAV motion control and synchronized sensor support for RGB, depth, LiDAR, pose, and IMU streams; Blender~\cite{blender} for scene-mesh conversion; Sionna RT~\cite{ait2025sionnart} for material-aware ray tracing; and CADFEKO~\cite{altairfeko} for UAV radar cross-section (RCS) modeling.
The resulting dataset contains 2.04 TB of data and 517,939 aligned multimodal data frames. 

LAMBDA is characterized by four key features.
First, it offers \textbf{high physical and visual fidelity}: the environment maps preserve detailed geometry and decimeter-level textures, refined optical and electromagnetic material information is assigned to scene objects, and UAV radar reflection characteristics are modeled using electromagnetic simulation. Second, it contains \textbf{diverse synchronized modalities}, including RGB images, depth maps, LiDAR point clouds, IMU states, CSI, and millimeter-wave radar-generation resources. Third, it covers \textbf{rich low-altitude scenarios}, including urban, suburban, open-ground, mountain, and campus scenes, multi-UAV/multi-base-station (multi-BS) settings, nighttime conditions, and rain, snow, and fog weather variations. Fourth, it provides \textbf{flexible RF and radar configuration support}. The stored path-level CSI records and radar-synthesis resources allow users to customize the antenna-array size, bandwidth, subcarrier spacing, and frequency-modulated continuous-wave (FMCW) chirp parameters for different sensing and communication tasks. They also support both plane-wave and spherical-wavefront channel synthesis, enabling far-field and near-field communication and sensing studies.

The reliability and usability of \dataset{} are validated through a multi-stage technical validation procedure. Generation-time quality control is first performed to check archive integrity, modality availability, timestamp alignment, pose continuity, and trajectory completeness. Cross-modal consistency is then examined using synchronized RGB, depth, LiDAR, pose, CSI, and radar visualizations, including weather-dependent RGB/depth/LiDAR comparisons, power-delay-profile inspection, and radar range--angle map sanity checks. Finally, two representative UAV ISAC tasks, namely RGB-aided beam prediction and RGB--LiDAR-based localization, are used to verify that the released data can be directly consumed by learning-based ISAC pipelines. These use cases are intended to demonstrate data usability and physical consistency rather than to establish task-specific state-of-the-art algorithms.

\section{Methods}
\label{sec:methods}
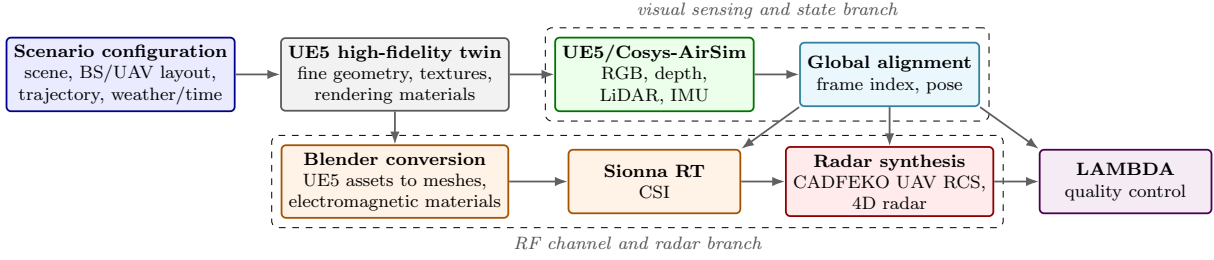
\begin{figure}[tb!]
\centering
\resizebox{0.98\textwidth}{!}{
\begin{tikzpicture}[
    font=\scriptsize,
    node distance=7mm and 7mm,
    >={Latex[length=2.0mm]},
    block/.style={
        draw,
        thick,
        rounded corners=2pt,
        align=center,
        inner sep=3.2pt,
        minimum width=27mm,
        minimum height=10mm
    },
    config/.style={block, fill=blue!8, draw=blue!55!black},
    scene/.style={block, fill=gray!10, draw=gray!70!black},
    sense/.style={block, fill=green!8, draw=green!45!black},
    radio/.style={block, fill=orange!10, draw=orange!65!black},
    radar/.style={block, fill=red!8, draw=red!55!black},
    data/.style={block, fill=cyan!7, draw=cyan!55!black},
    release/.style={block, fill=violet!8, draw=violet!60!black},
    group/.style={draw, rounded corners=3pt, dashed, inner sep=4pt},
    arrow/.style={->, thick, draw=gray!75!black},
    note/.style={align=center, font=\scriptsize\itshape, text=gray!60!black}
]
\node[config] (cfg) {\textbf{Scenario configuration}\\scene, BS/UAV layout,\\trajectory, weather/time};
\node[scene, right=of cfg] (dt) {\textbf{UE5 high-fidelity twin}\\fine geometry, textures,\\rendering materials};
\node[sense, right=of dt] (airsim) {\textbf{UE5/Cosys-AirSim}\\RGB, depth,\\LiDAR, IMU};
\node[data, right=of airsim] (sync) {\textbf{Global alignment}\\frame index, pose};

\coordinate (bottomrow) at ($(dt.south)+(0,-11mm)$);
\node[radio] (emmap) at (dt |- bottomrow) {\textbf{Blender conversion}\\UE5 assets to meshes,\\electromagnetic materials};
\node[radio] (sionna) at (airsim |- bottomrow) {\textbf{Sionna RT}\\CSI};
\node[radar] (radarsim) at (sync |- bottomrow) {\textbf{Radar synthesis}\\CADFEKO UAV RCS,\\4D radar};
\node[release, right=of radarsim] (release) {\textbf{LAMBDA}\\quality control};

\node[group, fit=(airsim)(sync), label={[note]above:visual sensing and state branch}] {};
\node[group, fit=(emmap)(sionna)(radarsim), label={[note]below:RF channel and radar branch}] {};

\draw[arrow] (cfg.east) -- (dt.west);
\draw[arrow] (dt.east) -- (airsim.west);
\draw[arrow] (airsim.east) -- (sync.west);
\draw[arrow] (dt.south) -- (emmap.north);
\draw[arrow] (emmap.east) -- (sionna.west);
\draw[arrow] (sync.south west) -- (sionna.north east);
\draw[arrow] (sionna.east) -- (radarsim.west);
\draw[arrow] (sync.south) -- (radarsim.north);
\draw[arrow] (radarsim.east) -- (release.west);
\draw[arrow] (sync.south east) -- (release.north west);
\end{tikzpicture}
}
\caption{Overview of the \dataset{} digital-twin generation pipeline.}
\label{fig:pipeline}
\end{figure}

To generate LAMBDA, we developed a digital-twin pipeline as shown in Fig.~\ref{fig:pipeline}.
For each scene, the scenario configuration is specified according to the building layout, including BS positions and orientations, UAV initial positions, UAV motion patterns, and weather/time settings.
These configurations are applied to high-fidelity UE5 digital-twin scenes that include detailed geometry, textures, rendering materials, lighting, and weather effects.
The configured scenes are then processed through two coupled generation branches.
In the visual sensing and state branch, Cosys-AirSim executes the configured UAV trajectory inside the UE5 scene and records RGB images, depth maps, LiDAR point clouds, IMU states, and ground-truth poses.
In the RF channel and radar branch, the same UE5 assets are exported through Blender, converted into radio-scene meshes, assigned electromagnetic material properties, and imported into Sionna RT for ray tracing and CSI generation.
The radar synthesis module further combines the Sionna RT multipath geometry with CADFEKO-derived UAV RCS information to generate radar data.
The outputs from both branches are finally passed to the LAMBDA quality-control module, which checks modality completeness, frame-index consistency, pose continuity, and CSI/radar plausibility.

The multimodal spatiotemporal alignment is performed offline, rather than by running all simulators simultaneously. Specifically, UE5/Cosys-AirSim first executes the UAV trajectory, renders the perception modalities, and records frame-indexed UAV and sensor poses.
All scene geometry, UAV poses, sensor poses, CSI records, and radar-synthesis metadata are represented in a unified Sionna right-handed world coordinate system. During generation, UE5/Cosys-AirSim coordinates are converted to this common coordinate convention through a left-handed-to-right-handed coordinate transformation.
Consequently, records with the same frame index share the same scene geometry, UAV state, timestamp, and coordinate convention across modalities.

\begin{figure}[tb!]
\centering
\setlength{\tabcolsep}{3pt}
\renewcommand{\arraystretch}{0.95}
\begin{tabular}{ccc}
\includegraphics[width=0.318\textwidth]{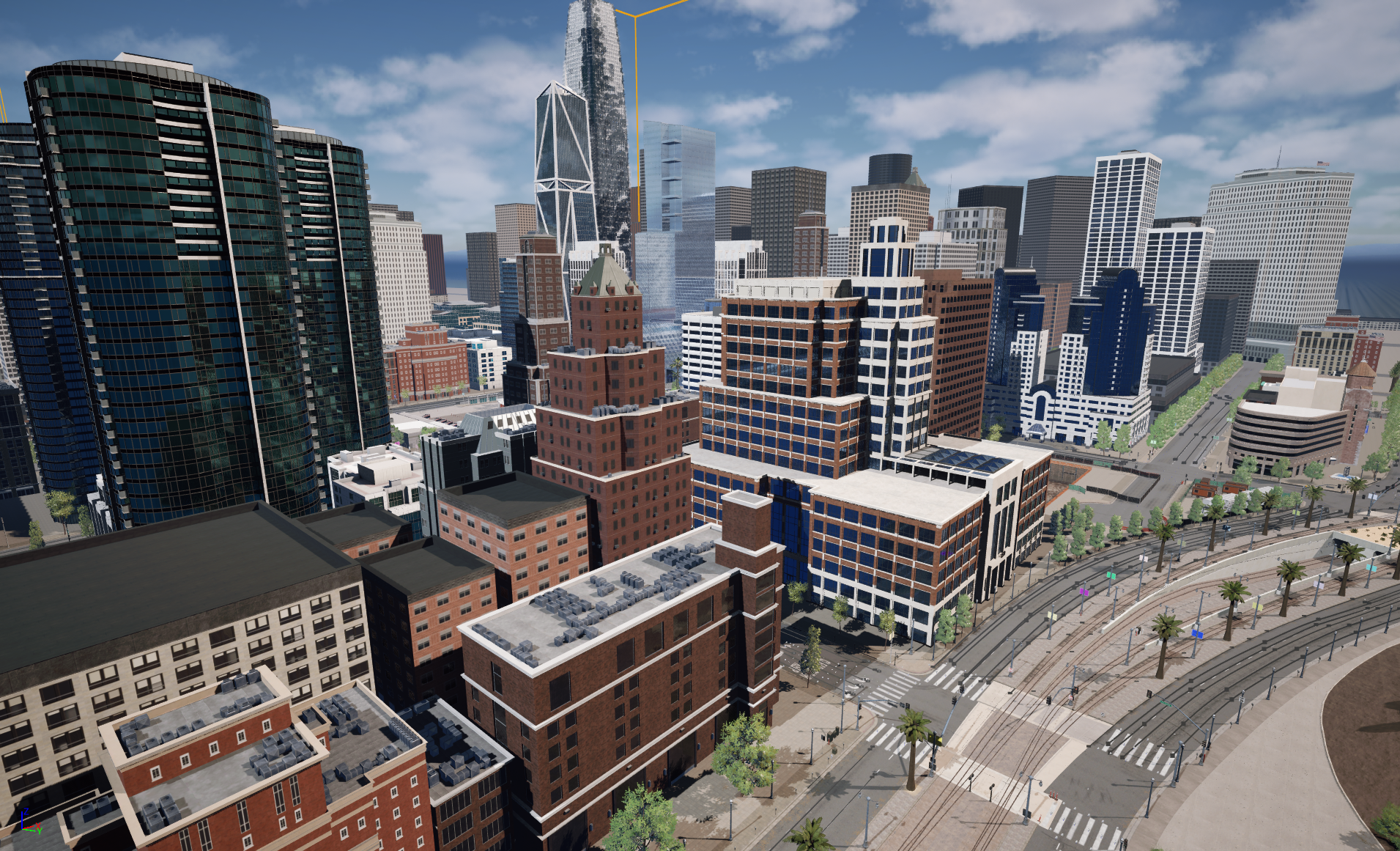} &
\includegraphics[width=0.318\textwidth]{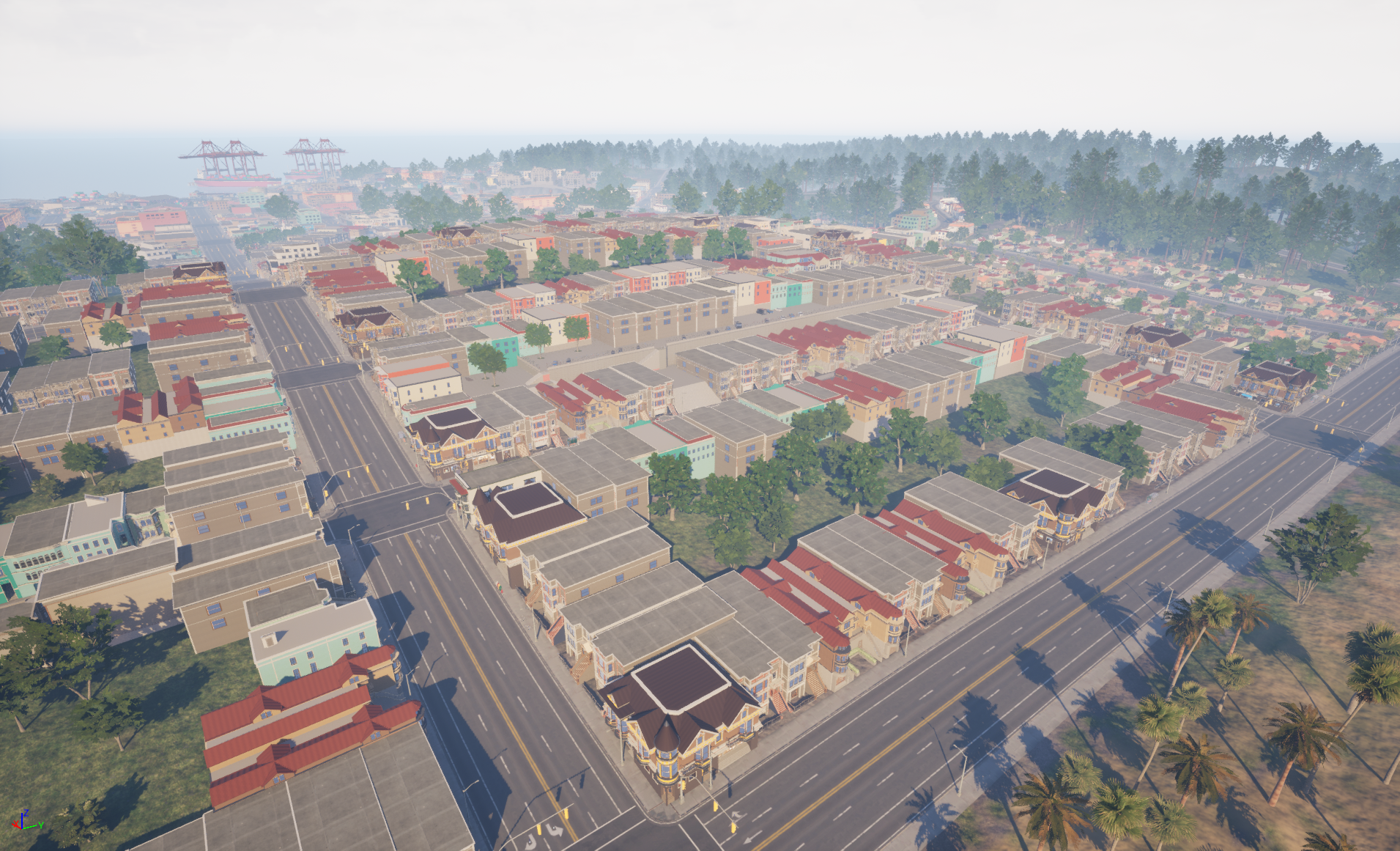} &
\includegraphics[width=0.318\textwidth]{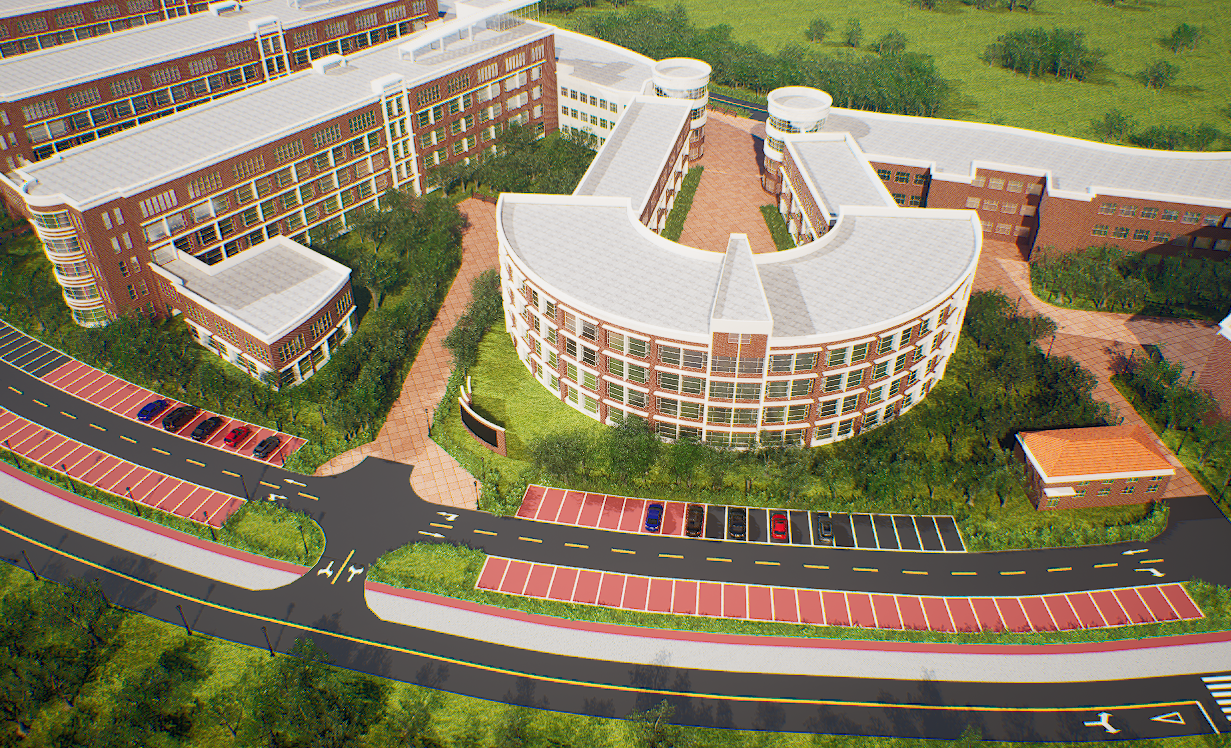} \\
\scriptsize (a) Urban Area &
\scriptsize (b) Suburbs &
\scriptsize (c) Campus
\end{tabular}
\caption{Regions included in the \dataset{} dataset.}
\label{fig:scene_gallery}
\end{figure}

\parahead{High-Fidelity Scene Construction} 
\dataset{} uses digital-twin scenes with detailed geometry, decimeter-level visual textures, and layered material assignments.
The dataset contains three main region categories, with representative renderings shown in Fig.~\ref{fig:scene_gallery} and scene-level parameters summarized in Table~\ref{tab:scene_parameters}.
The table reports the scene size, material diversity, UAV altitude range, BS--UAV configurations, weather setup, and number of frames for each scene.
The Urban region is based on San Francisco-style geometry, including dense skyscrapers, medium-density blocks, and low-density squares.
The Suburbs region covers lower-density suburban, mountain, and open-ground scenes with vegetation, rough terrain, and winding roads.
The Campus region represents part of the Shanghai Jiao Tong University (SJTU) campus around the School of Electronic Information and Electrical Engineering (SEIEE), comprising five buildings, parking lots, and surrounding lawns.

\begin{table}[tb!]
  \centering
  \caption{Summary of LAMBDA scenes, including scene footprint, material diversity, UAV altitude range, BS--UAV configurations, weather setup, and the total number of frames per scene. The frame count aggregates all listed BS--UAV and weather/time configurations for each scene.}
  \label{tab:scene_parameters}
  \scriptsize
  \setlength{\tabcolsep}{1.7pt}
  \renewcommand{\arraystretch}{1.12}
  \newcommand{\tablecell}[1]{\begin{tabular}[c]{@{}l@{}}#1\end{tabular}}
  \newcommand{\numcell}[1]{\begin{tabular}[c]{@{}r@{}}#1\end{tabular}}
  \newcommand{\headercell}[1]{\makebox[\linewidth][c]{#1}}
  \newcommand{\tallcell}[1]{\raisebox{-0.55\height}{#1}}
  \newcommand{\lowertallcell}[1]{\raisebox{-0.85\height}{#1}}
  \newcommand{\tallnumcell}[1]{\raisebox{-0.55\height}{#1}}
  \newcommand{\centercell}[1]{\begin{tabular}[c]{@{}c@{}}#1\end{tabular}}
  \newcommand{\bscell}[1]{\begin{tabular}[c]{@{}l@{}}#1\end{tabular}}
  \newcommand{\dropcell}[2][0.35ex]{\raisebox{-#1}{#2}}
  \begin{tabular}{@{}>{\raggedright\arraybackslash}m{0.115\textwidth}
                  >{\raggedright\arraybackslash}m{0.105\textwidth}
                  >{\raggedright\arraybackslash}m{0.125\textwidth}
                  >{\centering\arraybackslash}m{0.1\textwidth}
                  >{\centering\arraybackslash}m{0.075\textwidth}
                  >{\centering\arraybackslash}m{0.095\textwidth}
                  >{\centering\arraybackslash}m{0.105\textwidth}
                  >{\centering\arraybackslash}m{0.105\textwidth}
                  >{\raggedleft\arraybackslash}m{0.075\textwidth}@{}}
  \toprule
  \headercell{Region} & \headercell{Scenario} & \headercell{Scene} & \headercell{\shortstack{Footprint \\($\mathrm{m}\times\mathrm{m}$)}} & \headercell{\shortstack{Material \\types}} & \headercell{\shortstack{UAV Alt. \\range (m)}} & \headercell{\shortstack{BS--UAV \\setup}} & \headercell{\shortstack{Weather\\setup}} & \headercell{\shortstack{No. of\\frames}} \\
  \midrule
  \multirow{6}{=}[-0.9ex]{\mbox{Urban Area}} & \multirow{2}{=}[-1.5ex]{Block} & \tallcell{\texttt{Block\_1}} & \tallcell{249 $\times$ 192} & \tallcell{4} & \tallcell{46--129} & \tablecell{1 BS/1 UAV;\\2 BS/2 UAV} & \centercell{Sunny; rainy;\\snowy; foggy} & \tallnumcell{190,331} \\
  & & \texttt{Block\_2} & 340 $\times$ 221 & 5 & 50--123 & 1 BS/1 UAV & Sunny & \numcell{33,608} \\
  & \multirow{3}{=}[0ex]{Square} & \texttt{Square\_1} & 759 $\times$ 453 & 7 & 50--123 & 1 BS/1 UAV & Sunny & \numcell{33,608} \\
  & & \texttt{Square\_2} & 300 $\times$ 224 & 5 & 40--112 & 1 BS/1 UAV & Sunny & \numcell{33,608} \\
  & & \texttt{Square\_3} & 482 $\times$ 250 & 6 & 51--123 & 1 BS/1 UAV & Sunny & \numcell{33,608} \\
  & Skyscraper & \texttt{Skyscraper\_1} & 534 $\times$ 425 & 4 & 41--85 & 1 BS/1 UAV & Sunny & \numcell{36,060} \\
  \midrule
  \multirow{3}{=}[-4ex]{Suburbs} & \dropcell[0.55ex]{\tallcell{Suburb}} & \dropcell[0.8ex]{\tallcell{\texttt{Suburb\_1}}} & \dropcell[0.8ex]{\tallcell{265 $\times$ 226}} & \dropcell[0.8ex]{\tallcell{6}} & \dropcell[0.8ex]{\tallcell{41--71}} & \dropcell[2ex]{\bscell{1 BS/1 UAV;\\1 BS/2 UAV}} & \centercell{Sunny; night;\\rainy; snowy;\\foggy} & \dropcell[0.8ex]{\tallnumcell{112,261}} \\
  & Mountain & \tablecell{\texttt{Mountain\_1}} & \tablecell{1433 $\times$ 1237} & \tablecell{3} & \tablecell{62--96} & \bscell{1 BS/2 UAV} & \tablecell{Sunny; night} & \numcell{24,040} \\
  & Open ground & \texttt{Open\_Ground} & 63 $\times$ 45 & 3 & 23--55 & 1 BS/1 UAV & Sunny & \numcell{8,795} \\
  \midrule
  Campus & SJTU & \texttt{SEIEE} & 355 $\times$ 271 & 7 & 16--49 & 1 BS/2 UAV & Sunny & \numcell{12,020} \\
  \bottomrule
  \end{tabular}
\end{table}

The scene design emphasizes rooftop and infrastructure-mounted viewpoints looking into low-altitude airspace, where UAVs appear small and challenging to detect in visual, radar, and LiDAR modalities.
This setting reflects practical low-altitude monitoring scenarios, in which BSs or roadside units observe UAVs at distances ranging from tens to hundreds of meters.
The same digital-twin scenes are also reused for electromagnetic simulation. UE5 scene assets are exported through Blender, assigned customized electromagnetic material properties, and then imported into Sionna RT.
The customized electromagnetic material set covers common low-altitude scene surfaces, including concrete, brick, glass, metal, marble, wood, vegetation, and dry, wet, or snow-covered ground.

\begin{figure}[tb!]
\centering
{\setlength{\tabcolsep}{2pt}
\renewcommand{\arraystretch}{0.95}
\begin{tabular}{cc}
\includegraphics[width=0.43\columnwidth,height=0.27\columnwidth,keepaspectratio]{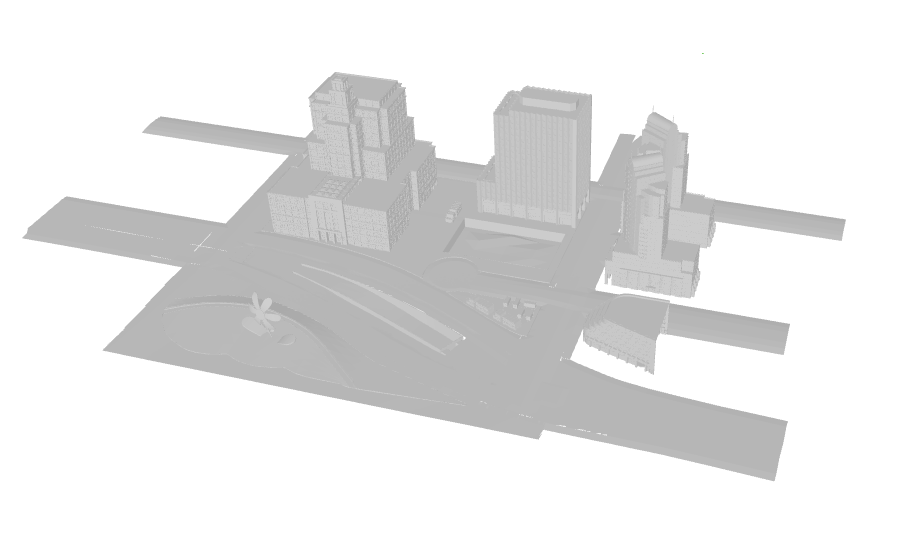} &
\includegraphics[width=0.43\columnwidth,height=0.27\columnwidth,keepaspectratio]{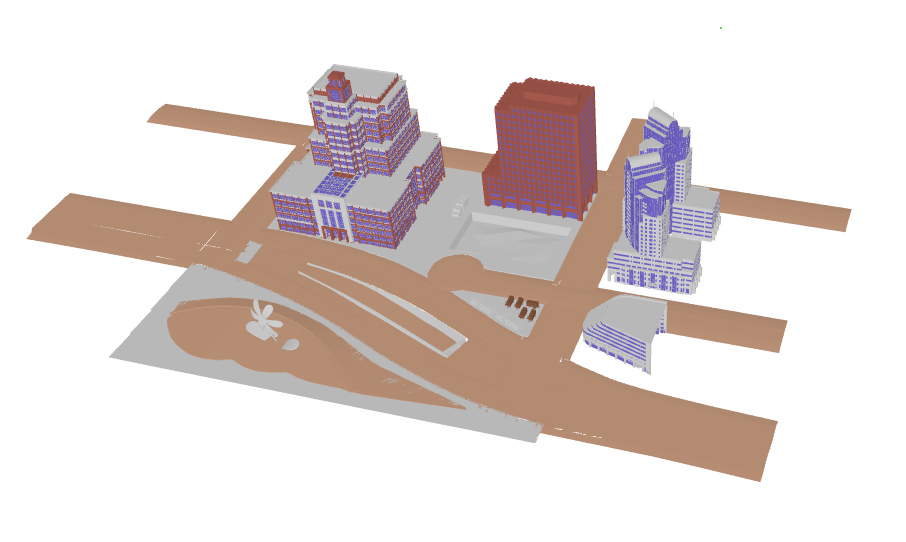} \\
\scriptsize (a) Uniform concrete baseline &
\scriptsize (b) Refined material assignment
\vspace{3mm}
\end{tabular}
\definecolor{matBrick}{HTML}{9B5850}
\definecolor{matGlass}{HTML}{7D74BF}
\definecolor{matConcrete}{HTML}{BABABA}
\definecolor{matDryGround}{HTML}{B28B72}
\definecolor{matWood}{HTML}{7A523C}
\begin{tikzpicture}[x=1cm,y=1cm,every node/.style={font=\scriptsize}]
\draw[draw=gray!35, line width=0.35pt, rounded corners=2pt] (0,0) rectangle (12.2,0.68);
\filldraw[fill=matBrick, draw=gray!45] (0.35,0.18) rectangle +(0.34,0.32);
\node[anchor=west] at (0.78,0.34) {Brick};
\filldraw[fill=matGlass, draw=gray!45] (2.15,0.18) rectangle +(0.34,0.32);
\node[anchor=west] at (2.58,0.34) {Glass};
\filldraw[fill=matConcrete, draw=gray!45] (3.95,0.18) rectangle +(0.34,0.32);
\node[anchor=west] at (4.38,0.34) {Concrete};
\filldraw[fill=matDryGround, draw=gray!45] (6.20,0.18) rectangle +(0.34,0.32);
\node[anchor=west] at (6.63,0.34) {Asphalt-like ground};
\filldraw[fill=matWood, draw=gray!45] (9.85,0.18) rectangle +(0.34,0.32);
\node[anchor=west] at (10.28,0.34) {Wood};
\end{tikzpicture}}
\caption{Material-assignment comparison for the \dataset{} \texttt{Square\_2} scene.}
\label{fig:material_comparison}
\end{figure}
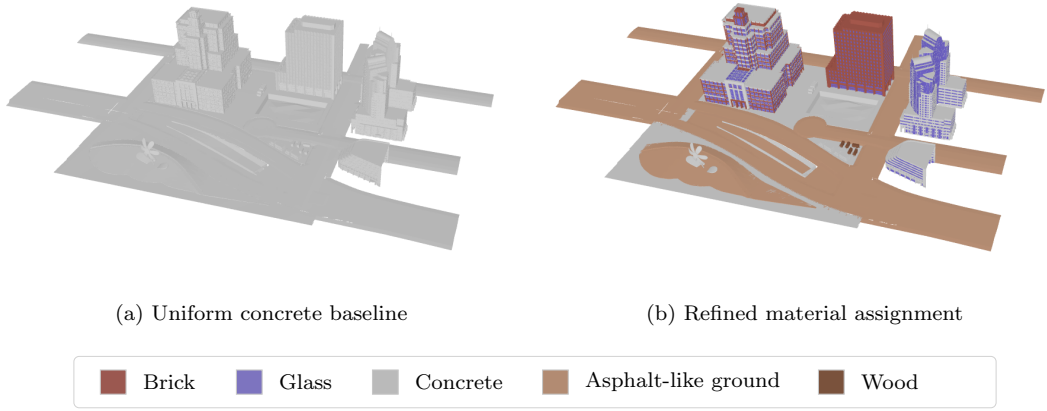

Fig.~\ref{fig:material_comparison} illustrates the visual difference between a uniform-material baseline and the refined material assignment used in \dataset{}.
The legend indicates the material-to-color mapping in the refined assignment; the asphalt-like ground corresponds to the customized \texttt{very\_dry\_ground} material used for road-like surfaces.
This comparison verifies that the semantic material labels are preserved in the rendered scene and can be consistently mapped to electromagnetic parameters for downstream simulation.

\begin{figure}[tb!]
\centering
{\setlength{\tabcolsep}{1pt}
\newcommand{\trajpanel}[1]{
\parbox[c][0.39\columnwidth][c]{0.43\columnwidth}{
\centering\includegraphics[width=0.43\columnwidth,height=0.39\columnwidth,keepaspectratio]{#1}}}
\begin{tabular}{cc}
\trajpanel{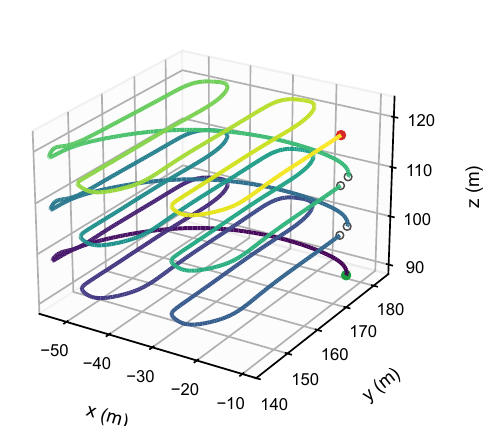} &
\trajpanel{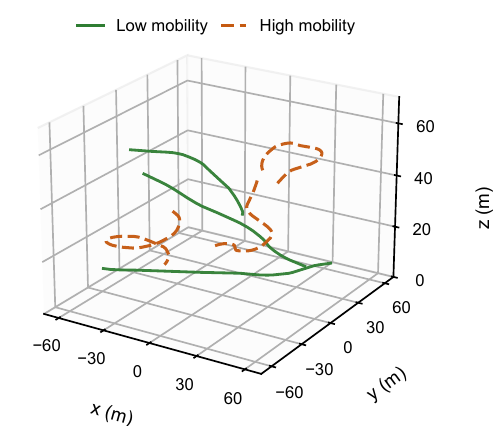} \\
\scriptsize (a) Realized z-traj segments &
\scriptsize (b) Random dynamic trajectories
\end{tabular}}
\caption{Example trajectories in \dataset.}
\label{fig:trajectory_examples}
\end{figure}

\parahead{Trajectory Control and Pose-based Alignment}
The spatiotemporal alignment of \dataset{} is built around frame-indexed UAV poses. 
During data collection, UAV trajectories are executed in Cosys-AirSim under a deterministic global simulation clock, and each global frame is assigned a timestamp and a realized UAV pose in the unified Sionna right-handed world coordinate system. 
The pose contains the UAV position and quaternion orientation at the corresponding simulation instant. 
Using the realized pose, rather than the idealized trajectory command, avoids mismatches caused by controller dynamics and numerical effects in AirSim. 
This pose-centered design makes the trajectory record the common alignment backbone for all sensing, motion, wireless, and radar-synthesis modalities. 
RGB, depth, LiDAR, IMU, and sensor-pose records are collected at the corresponding simulation frame, while CSI and radar products are generated offline from the same frame-wise UAV pose and converted scene geometry. 
Thus, the frame index acts as the primary key that links visual, geometric, inertial, wireless, and radar-synthesis records describing the same UAV state in the same world coordinate system.

The simulation adopts a global frame interval of \(1/60~\mathrm{s}\) for temporal alignment. 
In the data-collection loop, the simulator is paused before each acquisition so that all recorded sensor streams correspond to the same instantaneous scene state. 
RGB, depth, pose, and IMU records follow this 60 Hz global frame clock, while the 20 Hz LiDAR stream is recorded every three global frames and assigned the corresponding global frame index. 
Each global frame is indexed by the prescribed simulation instant \(t_k=k\Delta t\), with \(\Delta t=1/60~\mathrm{s}\). 
The collector also records simulator timestamps to detect stale or duplicated sensor outputs.

For trajectory control, \dataset{} provides two types of UAV motion patterns: layered \texttt{z-traj} sweeps and random dynamic trajectories with low- and high-mobility presets. 
The \texttt{z-traj} sweep is a deterministic layer-wise trajectory designed to sample the same horizontal footprint at multiple UAV altitudes. 
For a square region with side length \(L\), the generator creates altitude layers spaced by \(\Delta_h\), places parallel scan lanes separated by \(\Delta_\ell\) within each layer, and connects adjacent lanes in a serpentine order with spatial step \(v\Delta t\). 
In \dataset{}, the \texttt{z-traj} sweeps are configured with \(L=50~\mathrm{m}\), \(\Delta_\ell=10~\mathrm{m}\), \(\Delta_h=10~\mathrm{m}\), and \(v=5~\mathrm{m/s}\). 
These records enable controlled altitude-dependent comparisons of visual, geometric, and wireless-channel behavior. 
Accordingly, they are treated as altitude-layer scan segments rather than as a single continuous 3D maneuver, as illustrated in Fig.~\ref{fig:trajectory_examples}(a).

The random dynamic trajectories are generated using a Markov-chain maneuvering model inspired by maneuvering-target dynamics~\cite{li2003maneuvering}. 
The model switches among constant-velocity, constant-acceleration, coordinated-turn, and Singer-style dynamics, with initial positions, speeds, headings, and vertical velocities sampled by Latin hypercube sampling to improve flight-volume coverage. 
The UAV state is then evolved frame by frame according to the sampled motion mode, process noise, and boundary-handling rules. 
Low- and high-mobility presets share the same generator but use different maneuverability limits: the acceleration components are clipped to \(\pm 4.0~\mathrm{m/s^2}\) and \(\pm 6.0~\mathrm{m/s^2}\), and the coordinated-turn-rate limits are \(80^\circ/\mathrm{s}\) and \(120^\circ/\mathrm{s}\), respectively. 
Both presets bound the horizontal speed by \(13.5~\mathrm{m/s}\) and the vertical speed by \(6.0~\mathrm{m/s}\). 
Representative low- and high-mobility trajectories are shown in Fig.~\ref{fig:trajectory_examples}(b).

\parahead{Wireless Channel Modeling} CSI-related channel records are generated using Sionna RT. Each record contains complex-valued path gains, delays, Doppler shifts, angles of departure and arrival, along with interaction labels, UAV position and velocity, carrier frequency, and weather kind. The Doppler shifts are determined by the UAV velocity projected onto each propagation path direction.
The CSI subsets cover multiple candidate carrier frequencies for low-altitude wireless networks and typical 5G and 6G bands, including 1.4, 3.5, 4.9, 5.9, 7, 28, 60, and 77 GHz. 
The CSI generation pipeline allows users to customize the subcarrier configuration and antenna arrays and generate the channel response of the corresponding orthogonal frequency-division multiplexing (OFDM) signals based on these path-level channel records. The pipeline further supports both plane-wave and spherical-wavefront channel synthesis because the released path-level records preserve geometric propagation parameters, rather than fixed antenna-domain channel tensors, allowing users to recompute array responses under either far-field or near-field assumptions.

At frequencies of 28 GHz, 60 GHz, and 77 GHz, the pipeline uses frequency-dependent radio-material and scattering settings together with ITU-based atmospheric attenuation for adverse-weather RF paths.
The pipeline remaps selected material parameters to support frequencies up to the millimeter-wave range and introduces frequency-dependent scattering coefficients for rough urban surfaces such as asphalt.
It also supports gaseous attenuation models such as oxygen and water-vapor absorption.
Detailed weather settings are implemented as a modular configuration component for weather-specific data subsets.

\parahead{Radar Signal Generation}
Radar signal generation is implemented as a configurable downstream component, rather than as a separately stored raw-radar modality.
Given the released path-level CSI records, UAV pose and attitude, radar configuration, and UAV RCS files, the provided code synthesizes dechirped FMCW radar cubes with antenna, chirp, and range-bin dimensions.
The synthesis combines the complex CSI coefficient of each valid path with its two-way delay, Doppler shift, virtual-array steering phase, UAV-orientation-dependent RCS, and optional receiver noise.
The complex-valued CSI coefficient captures the propagation effects computed by the ray-tracing channel generator, whereas the RCS term is obtained from CADFEKO simulations of the AirSim UAV model and queried according to the UAV attitude.
The radar-generation parameters are user-configurable, including the bandwidth, chirp duration, sampling rate, number of chirps per frame, receiver noise level, and antenna array shape and spacing.
As a reference configuration, the released code provides a 77 GHz preset with 2 GHz bandwidth, 40 \(\mu\)s chirp duration, a 204.8 MHz sampling rate, 64 chirps per frame, and a \(4\times4\) antenna array.
Users may modify these chirp and array parameters to synthesize radar cubes corresponding to different FMCW radar designs.
The generated radar cubes are stored after a fast Fourier transform (FFT) is applied along the fast-time dimension.
For visualization and validation, range--Doppler maps are formed from the stored range bins and Doppler FFTs over chirps, while range--azimuth and range--elevation maps are obtained by applying angle FFTs over the rectangular virtual array.

\parahead{Weather Configuration} 
Weather effects are implemented through modality-specific pipelines because RF propagation, image rendering, depth sensing, and LiDAR point-cloud formation are governed by different degradation mechanisms.
For RF data, the CSI generator applies atmospheric attenuation to the complex multipath amplitudes, and radar synthesis then uses the resulting multipath records together with UAV RCS and FMCW configuration.
For RGB, depth, and LiDAR streams, weather degradation is simulated with the UE5 Niagara particle system, the LISA algorithm~\cite{lisa}, the UE5 volumetric-fog model, and the physically based fog modeling method of Hahner et al.~\cite{hahner2021fog}.
Table~\ref{tab:weather_method_assignment} summarizes the method assignment for the weather-varied modalities.

\begin{table}[tb!]
\centering
\setlength{\abovecaptionskip}{1pt}
\setlength{\belowcaptionskip}{2pt}
\caption{Modality-specific weather-degradation simulation methods used in LAMBDA.}
\label{tab:weather_method_assignment}
\scriptsize
\setlength{\tabcolsep}{5pt}
\renewcommand{\arraystretch}{1.03}
\resizebox{\textwidth}{!}{
\begin{tabular}{llll}
\toprule
Modality & Rain & Snow & Fog \\
\midrule
RF channel 
& ITU-R P.838-3 attenuation 
& Gunn--East wet-snow path loss 
& ITU-R P.840-9 attenuation \\

RGB 
& Niagara particle system 
& Niagara particle system 
& UE5 volumetric fog \\

Depth 
& Niagara particle system 
& Niagara particle system 
& Physics-based fog modeling \\

LiDAR 
& LISA algorithm 
& LISA algorithm 
& Physics-based fog modeling \\
\bottomrule
\end{tabular}
}
\end{table}

\taskhead{\textbf{RF Weather Attenuation Modeling}}
For RF data, weather-dependent propagation loss is represented by frequency-dependent electromagnetic attenuation terms rather than by rendered or point-cloud artifacts.
The CSI generator applies these attenuation terms to the complex multipath amplitudes, and radar synthesis then uses the resulting multipath records together with UAV RCS and FMCW configuration.
Rain attenuation follows ITU-R P.838-3~\cite{itu838}: the horizontal and vertical polarization coefficients are evaluated from the carrier frequency and then projected using the elevation angle and polarization tilt.
Fog and cloud attenuation follows ITU-R P.840-9~\cite{itu840}, using liquid-water density and temperature-dependent dielectric parameters.
Snow attenuation for RF paths follows the Gunn--East snow path-loss model for RF frequencies, using propagation distance, carrier frequency, and snow-equivalent precipitation rate as inputs~\cite{gunn1954microwave}.

\taskhead{\textbf{Niagara Particle System}}
For rain and snow, RGB images and depth maps are rendered in UE5 using Niagara particle systems. 
The particle geometry, opacity, velocity, acceleration, spawn rate, and emitter extent jointly determine the precipitation appearance and the density of visible hydrometeors within the sensor field of view. 
To maintain camera-frustum coverage while reducing rendering cost, the rain and snow emitters are placed locally around the BS sensors rather than across the entire scene. 
In the released configuration, both weather conditions use \(10\times10\times10~\mathrm{m^3}\) local emitters, with weather-specific particle velocities, accelerations, and spawn rates. 
The corresponding Niagara settings are released with the weather-configuration files to support reproducible rain and snow rendering.

\taskhead{\textbf{LISA Algorithm}}
The LiDAR rain and snow point clouds are synthesized from clear-weather LiDAR scans using the LISA algorithm~\cite{lisa}.
For each laser ray, LISA samples precipitation particles along the beam path and compares their simulated backscatter with the return from the original scene surface.
When the sampled particle return is stronger, the measured point is replaced by a precipitation-induced return; otherwise, the original object return is retained with scattering-induced range perturbation.
This branch therefore models sparse spurious returns and range noise caused by discrete hydrometeors, while keeping the LiDAR output synchronized with the same trajectory and weather labels as the RGB/depth streams.

\taskhead{\textbf{UE5 Volumetric Fog}}
Fog is treated separately from rain and snow because it is a spatially continuous participating medium rather than a set of resolvable particles.
For RGB images, fog is rendered with the UE5 volumetric-fog model instead of Niagara particles.
The visual attenuation and in-scattered radiance are accumulated along camera rays by the renderer, which yields a continuous fog layer that is more consistent with atmospheric extinction than a representation based on individually rendered fog particles.

\taskhead{\textbf{Physics-based Fog Modeling}}

For depth and LiDAR, fog degradation follows the physics-based simulation method of Hahner et al.~\cite{hahner2021fog}. 
The meteorological optical range (MOR) determines the extinction coefficient \(\alpha=\ln(20)/\mathrm{MOR}\), while atmospheric backscatter is modeled using the inverse-visibility scaling in~\cite{hahner2021fog}. 
For each sensing ray, the simulator compares the attenuated hard-target return with the soft fog backscatter response, which is obtained from a precomputed lookup table accounting for the range-dependent fog profile, laser pulse shape, and near-field transmitter--receiver overlap. 
If the soft response dominates, the measured range is replaced by the corresponding fog-return range; otherwise, the original target range is retained. 
The same hard-target/soft-return decision model is applied image-wise to depth maps and point-wise to LiDAR scans, with replaced LiDAR points moved along their original rays.

\section{Data Records}
\label{sec:data-records}
\begin{figure}[!htbp]
\centering
\makebox[\textwidth][c]{\includegraphics[width=1.06\textwidth]{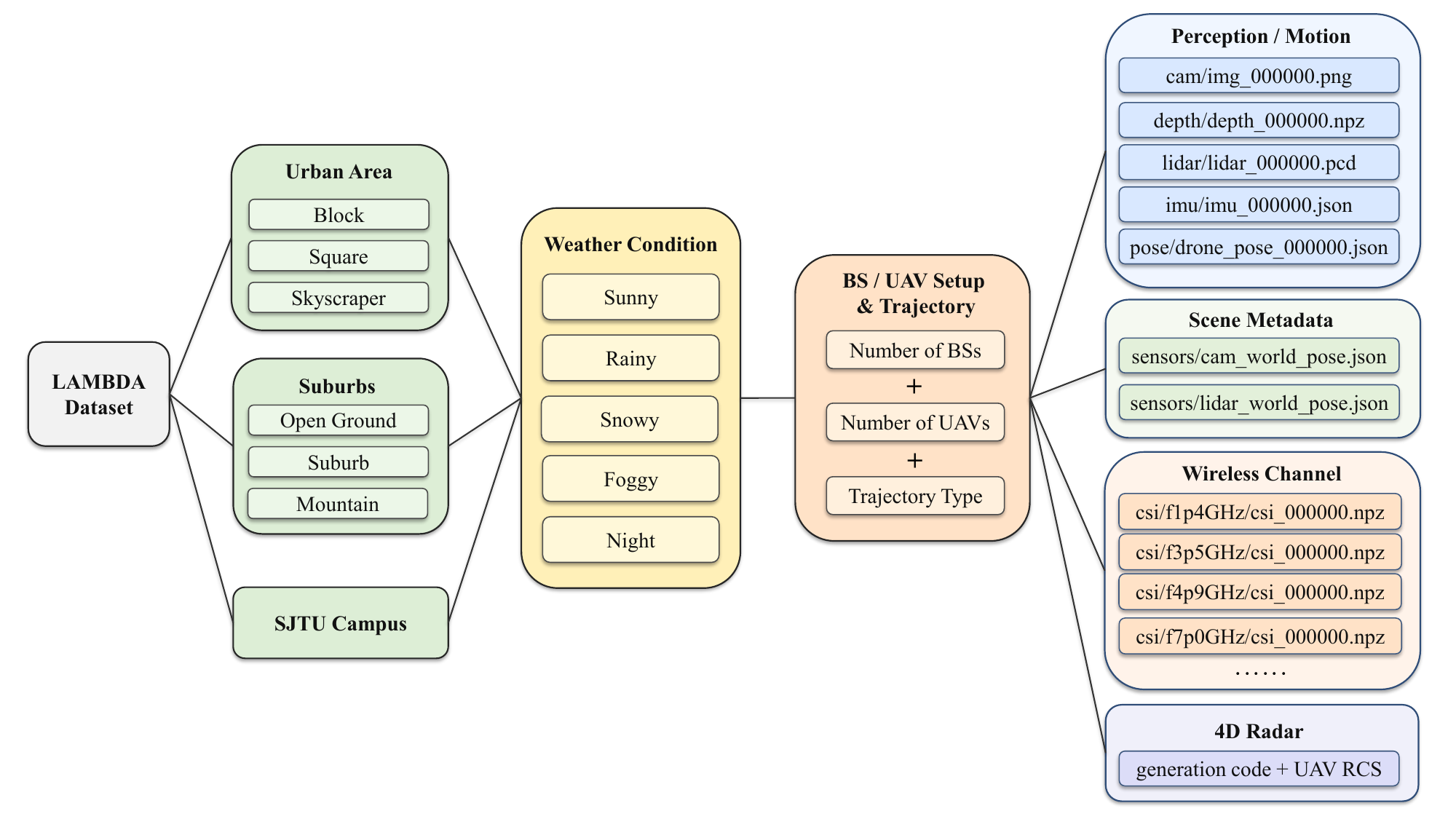}}
\caption{Data directory hierarchy and files of the \dataset{} dataset.}
\label{fig:file_structure}
\end{figure}

\dataset{} is organized as a synchronized low-altitude multimodal record set in which each global frame index is associated with visual, geometric, inertial, wireless channel, and radar information from the same UAV trajectory.
The whole dataset is deposited in Science Data Bank~\cite{zhou2026lambda_dataset}.
As shown in Fig.~\ref{fig:file_structure}, the dataset directory first separates scenes, and each scene is then organized by weather condition, BS--UAV setup and trajectory, and modality.
The dataset uses ZIP archives to reduce storage overhead while preserving the frame-indexed file structure: deterministic trajectory subsets are mainly packaged by modality, while random dynamic subsets are separated into \texttt{low\_mobility} and \texttt{high\_mobility} folders and compressed by individual trajectory as \texttt{traj\_*.zip} archives.

The RF data include BS-to-UAV CSI files and radar-generation resources.
The CSI files store multipath channel information, including complex-valued path gains, delays, Doppler shifts, angles of departure and arrival, path validity flags, interaction labels, UAV position and velocity, carrier frequency, and weather kind.
These path-level data are stored as compressed NumPy files under carrier-frequency-specific CSI archives, and can be used to generate CSI under configurable setups. 
For radar sensing, \dataset{} follows an FMCW radar model, where each chirp sweeps its carrier frequency over time. The target range, velocity, and angle can be estimated from the beat frequency, Doppler shift, and antenna-array phase.
Instead of storing raw radar cubes for every frame, \dataset{} provides radar-generation and visualization code together with UAV RCS files.
The provided code combines the stored CSI multipath parameters, the UAV RCS model, and the radar configuration to synthesize FMCW radar data, from which range-Doppler and range-angle maps can be generated for visualization and analysis.

The non-RF sensing data include RGB images, depth maps, LiDAR point clouds, IMU states, UAV poses, and BS sensor poses.
RGB images are rendered from BS viewpoints and stored as PNG files.
Depth maps are pixel-aligned with the RGB frames and stored under \texttt{depth/} as compressed NumPy files.
LiDAR point clouds are stored as PCD files, while IMU and pose streams are stored as JSON files.
BS camera and LiDAR poses are saved once under the \texttt{sensors/} folder.
Representative sensor and wireless configurations are summarized in Table~\ref{tab:sensor_config}.

\begin{table}[tb!]
\centering
\caption{Representative sensor and wireless configurations.}
\label{tab:sensor_config}
\scriptsize
\setlength{\tabcolsep}{4pt}
\resizebox{\columnwidth}{!}{
\begin{tabular}{cll}
\toprule
Modality & Configuration & Values \\
\midrule
\multirow{1}{*}{CSI} & Carrier frequency & 1.4, 3.5, 4.9, 5.9, 7, 28, 60, and 77 GHz \\
\cmidrule(lr){1-3}
\addlinespace[1pt]
RGB & Field of view (FOV) / frame rate & \(100^\circ\) horizontal, \(63^\circ\) vertical, 60 Hz \\
\addlinespace[1pt]
\cmidrule(lr){1-3}
Depth & FOV / frame rate & \(100^\circ\) horizontal, \(63^\circ\) vertical, 60 Hz \\
\addlinespace[1pt]
\cmidrule(lr){1-3}
\addlinespace[1pt]
\multirow{2}{*}{LiDAR} & FOV / heading / & \(120^\circ\) horizontal, \(90^\circ\) vertical, camera-aligned orientation \\
& Channels / range / scan rate & 512 lines, 150 m, 20 Hz \\
\addlinespace[1pt]
\cmidrule(lr){1-3}
\addlinespace[1pt]
\multirow{2}{*}{Radar} & Antenna / heading / & \(4\times4\) array, camera-aligned heading \\
& FMCW settings & 77 GHz carrier, 2 GHz bandwidth, 40 \(\mu\)s chirp duration, 64 chirps/frame \\
\bottomrule
\end{tabular}
}
\end{table}

Table~\ref{tab:data_records_schema} summarizes the representative file organization of \dataset{}. Frame-indexed modality files share the same six-digit index whenever the corresponding modality is available.

\begin{table}[tb!]
\centering
\caption{Representative file organization for synchronized \dataset{} samples.}
\label{tab:data_records_schema}
\small
\setlength{\tabcolsep}{3.5pt}
\begin{tabular}{@{}>{\raggedright\arraybackslash}p{0.17\textwidth}>{\raggedright\arraybackslash}p{0.27\textwidth}>{\raggedright\arraybackslash}p{0.50\textwidth}@{}}
\toprule
Directory & Typical content & Example file / key fields\\
\midrule
\texttt{csi/} & BS-to-UAV multipath channel records & \texttt{csi\_000000.npz}, path gain, delay, angle of arrival/departure \\
\texttt{cam/} & BS-view RGB images & \texttt{img\_000000.png} \\
\texttt{depth/} & BS-view depth maps & \texttt{depth\_000000.npz} \\
\texttt{lidar/} & Point clouds & \texttt{lidar\_000000.pcd}, \(x,y,z\) \\
\texttt{imu/} & UAV inertial states & \texttt{imu\_000000.json}, linear acceleration, angular velocity, orientation \\
\texttt{pose/} & UAV pose & \texttt{drone\_pose\_000000.json}, position, orientation, timestamp \\
\texttt{sensors/} & BS sensor pose & \texttt{cam\_world\_pose.json}, \texttt{lidar\_world\_pose.json}, extrinsics \\
\bottomrule
\end{tabular}
\end{table}

\section{Technical Validation}
\label{sec:technical-validation}

The reliability and usability of \dataset{} are evaluated from complementary perspectives, including generation-time quality control, frame-level consistency verification, synchronized multimodal visualization, and two UAV ISAC use cases: RGB-aided beam prediction and RGB--LiDAR-based UAV localization.

\parahead{Quality Control and Consistency Checks}  \dataset{} adopts a multi-stage validation protocol to ensure that the released records are complete, synchronized, and physically plausible. During data generation, visual playback is used to inspect scene rendering, UAV motion, sensor viewpoints, and weather effects. Frame-level timestamp tests and pose-continuity checks are then applied to detect stale sensor outputs, duplicated frames, missing modality files, and discontinuous trajectories. For the RF and radar branches, ray-tracing path visualization, power-delay-profile inspection, and radar-map sanity checks are performed to verify channel-path plausibility and radar-synthesis consistency. 

\begin{figure}[tb!]
\centering
{\setlength{\tabcolsep}{2pt}
\renewcommand{\arraystretch}{0.95}
\newcommand{\weatherpanel}[1]{
\parbox[c][0.17\textwidth][c]{0.318\textwidth}{
\centering\includegraphics[width=0.318\textwidth,height=0.17\textwidth,keepaspectratio]{#1}}}
\begin{tabular}{ccc}
\weatherpanel{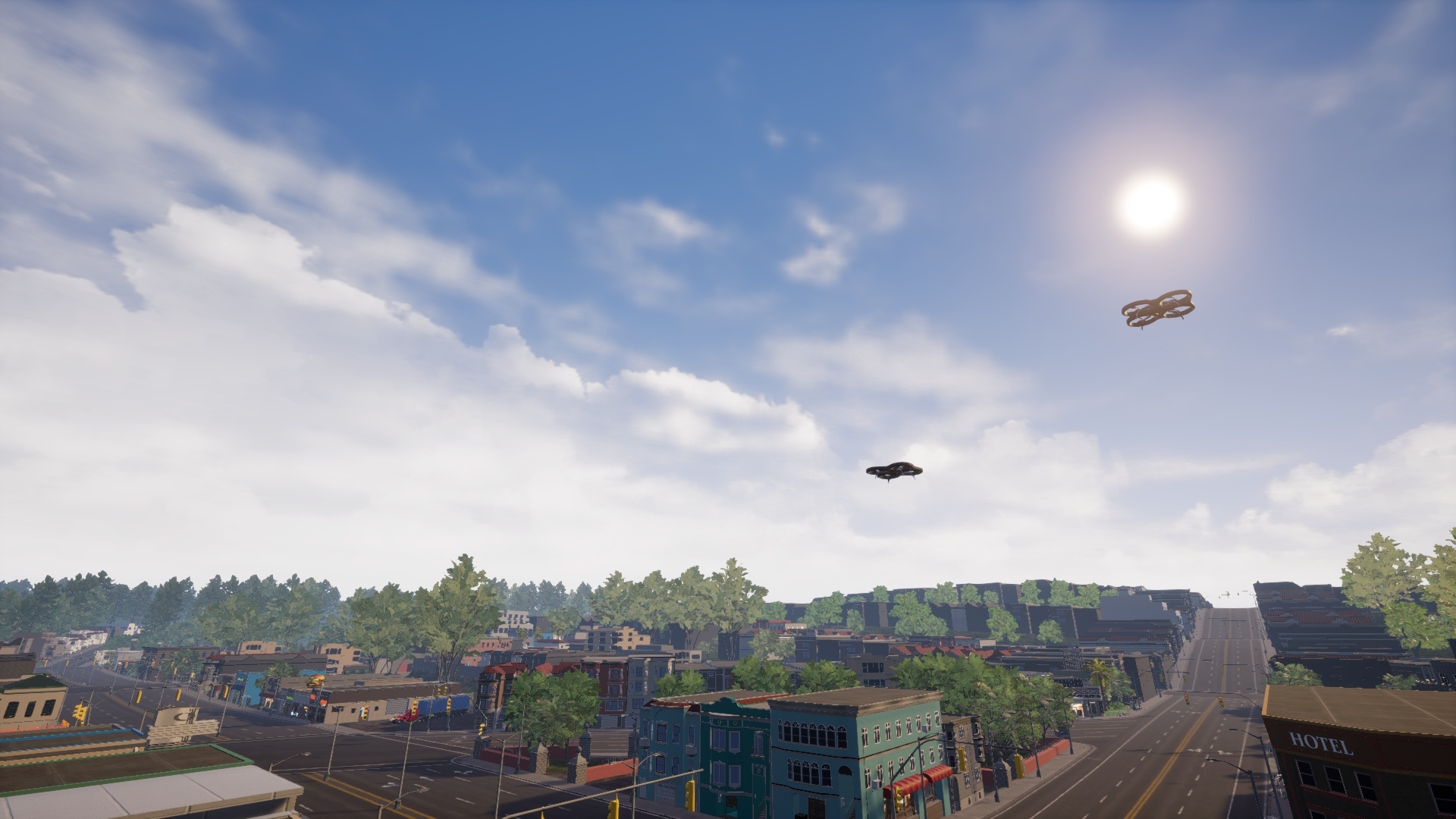} &
\weatherpanel{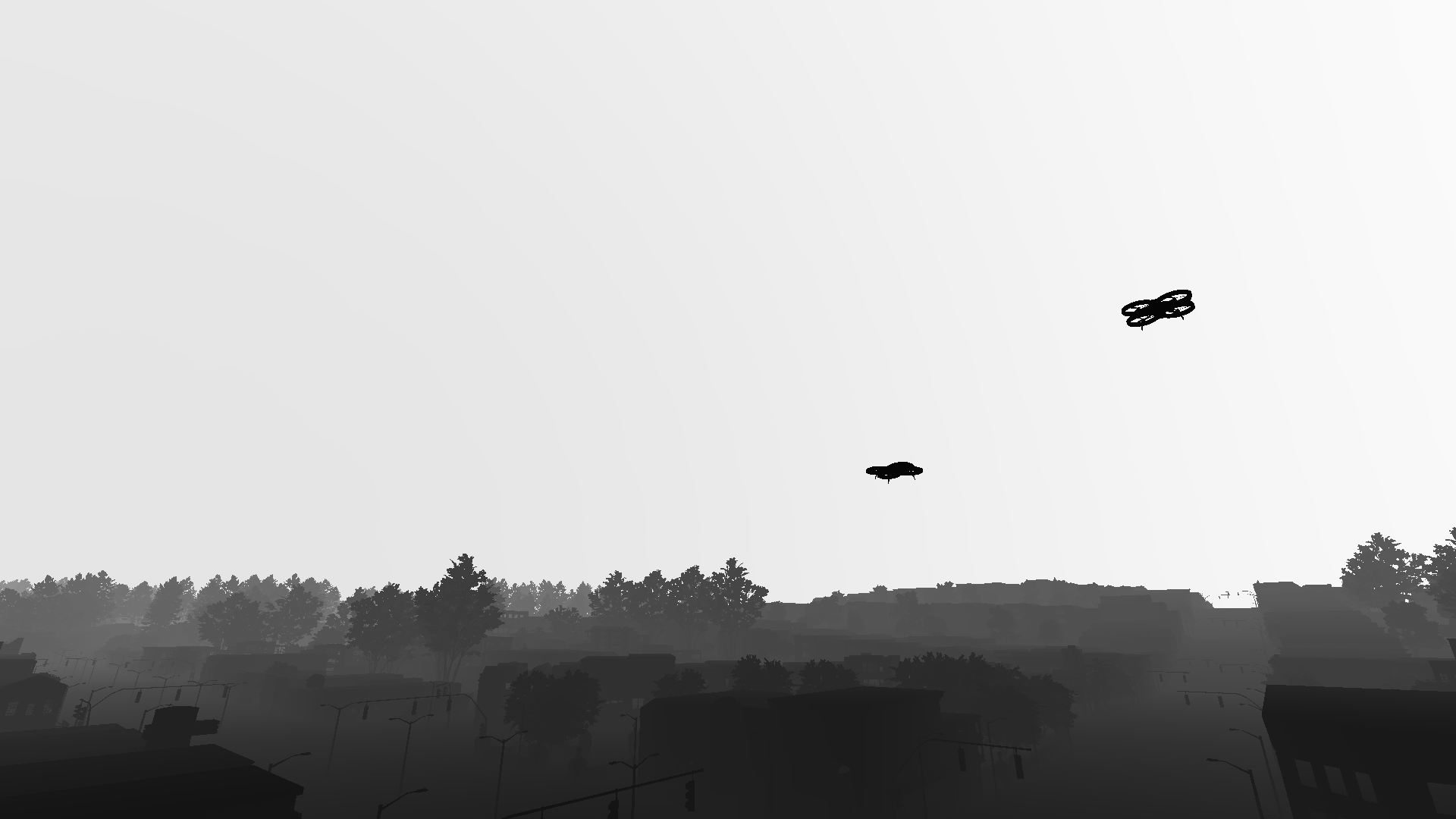} &
\weatherpanel{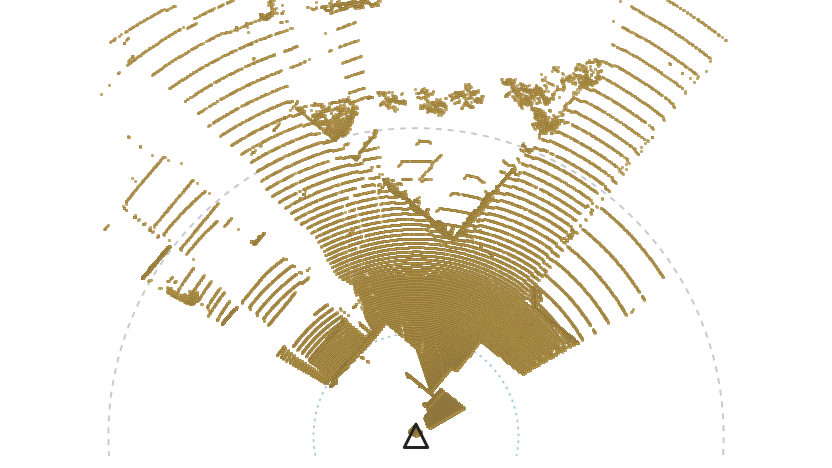} \\
\scriptsize (a) RGB, sunny &
\scriptsize (b) depth, sunny &
\scriptsize (c) LiDAR, sunny \\[0.5mm]
\weatherpanel{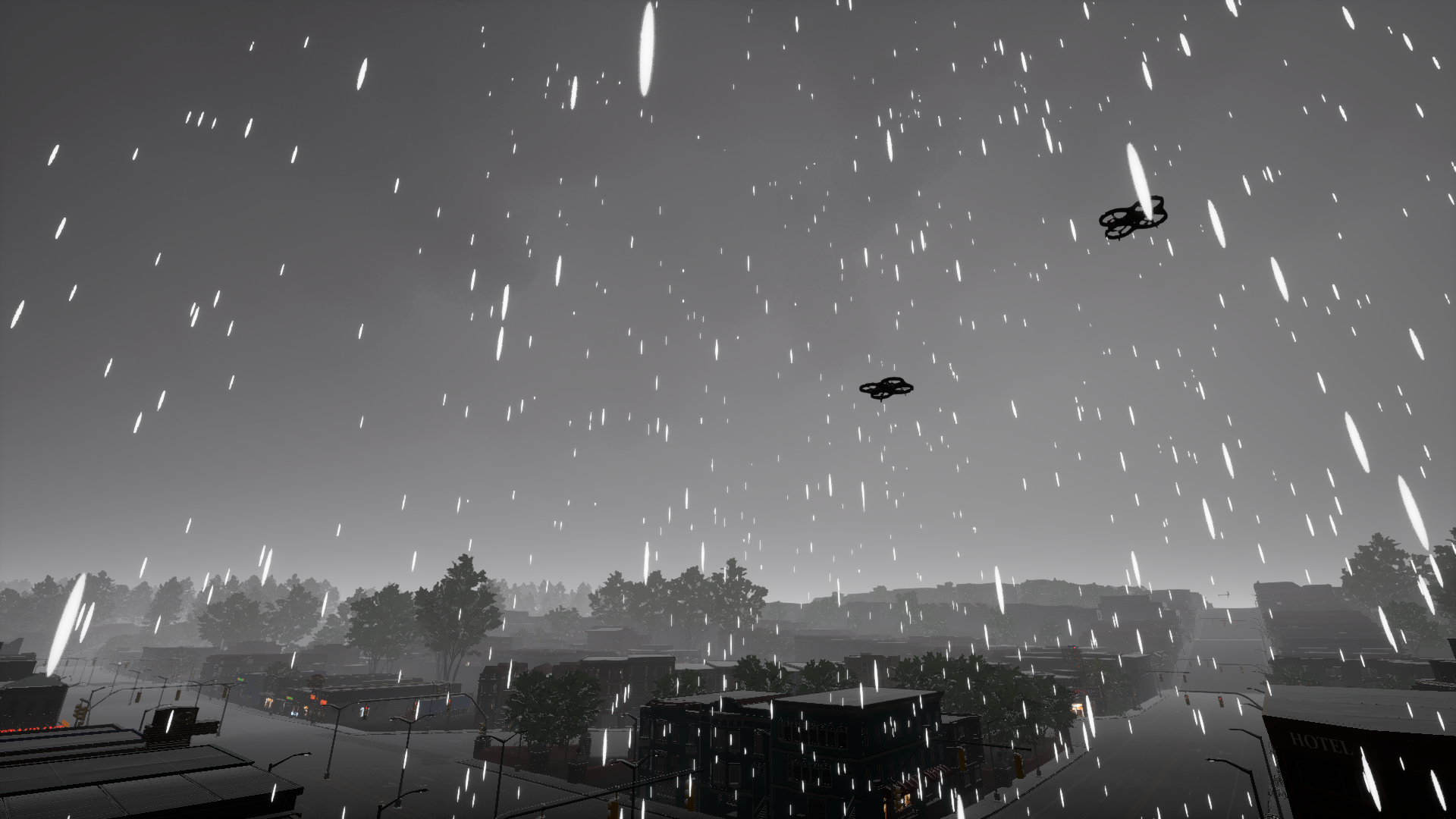} &
\weatherpanel{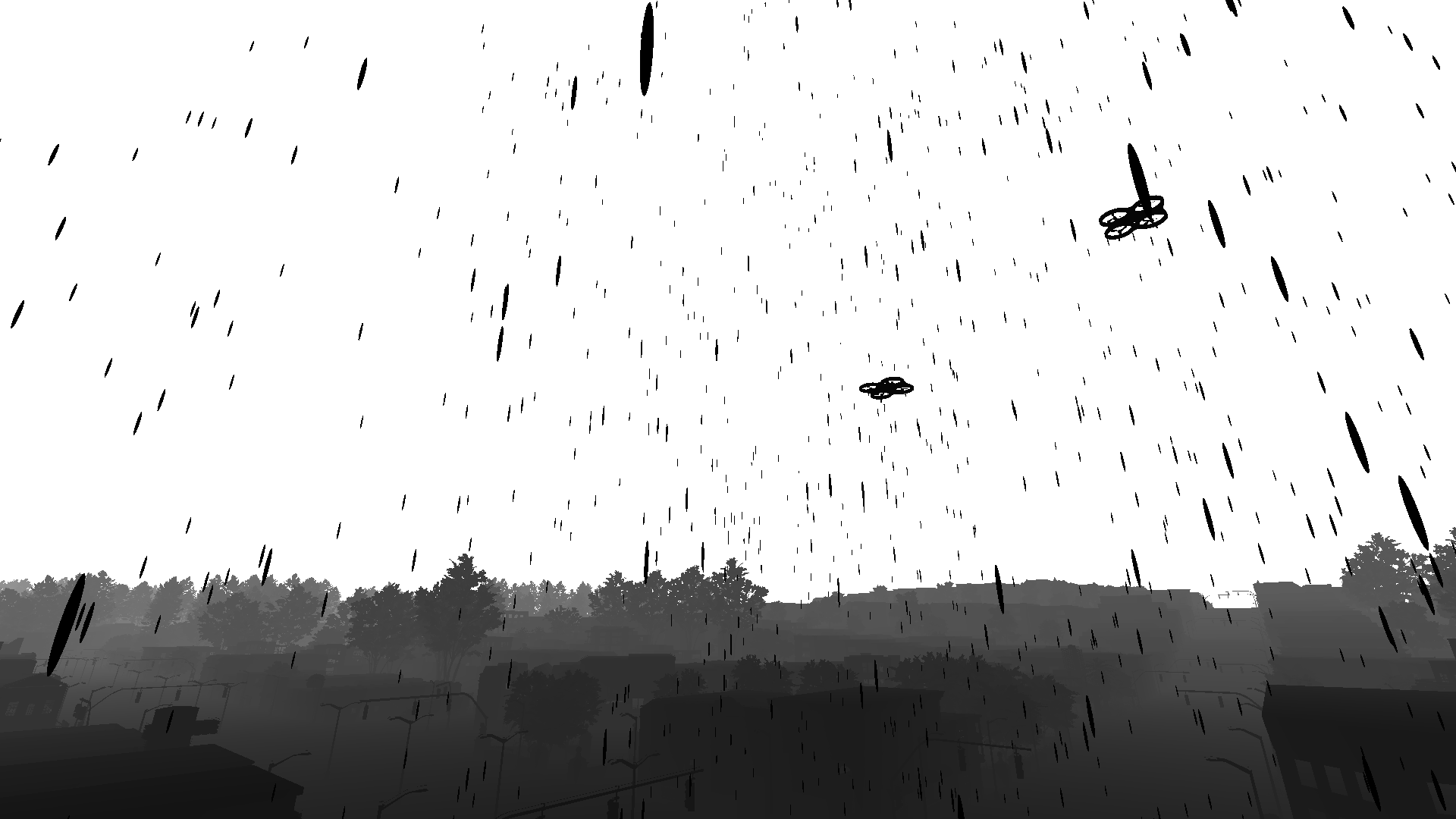} &
\weatherpanel{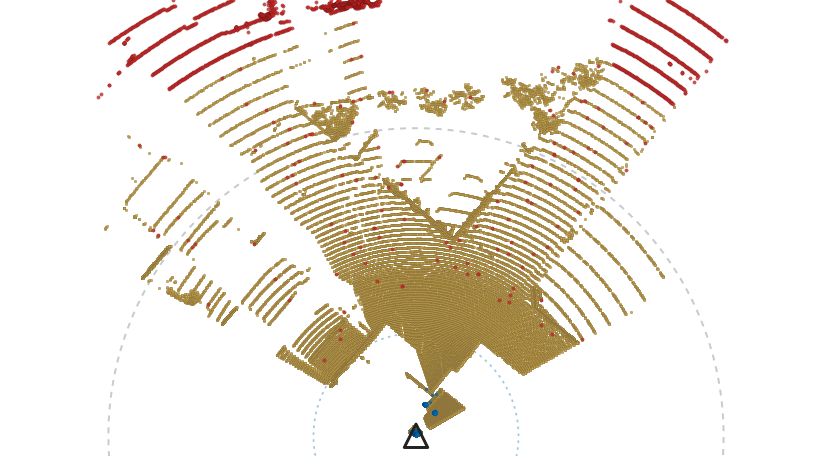} \\
\scriptsize (d) RGB, rainy &
\scriptsize (e) depth, rainy &
\scriptsize (f) LiDAR, rainy \\[0.5mm]
\weatherpanel{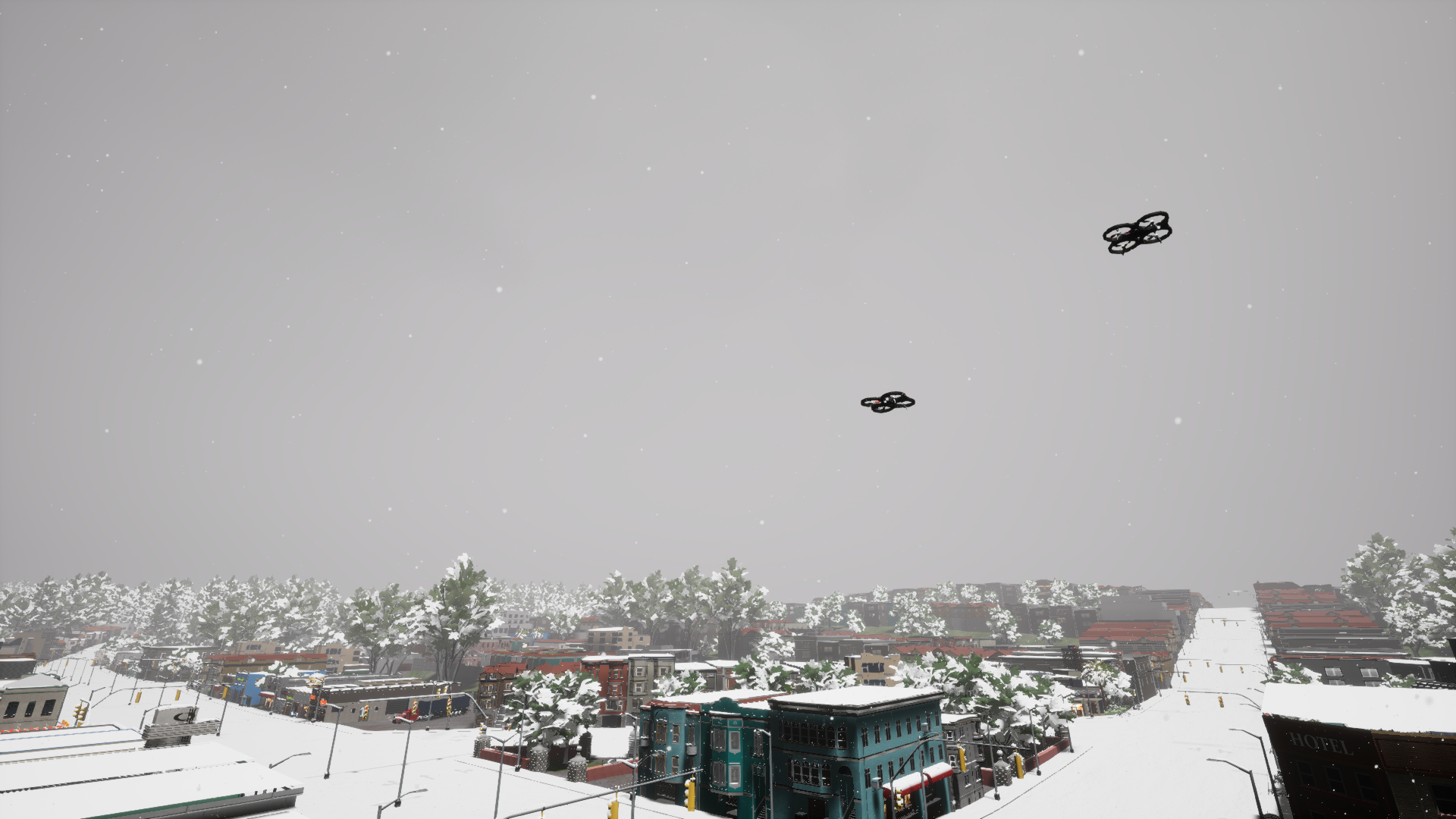} &
\weatherpanel{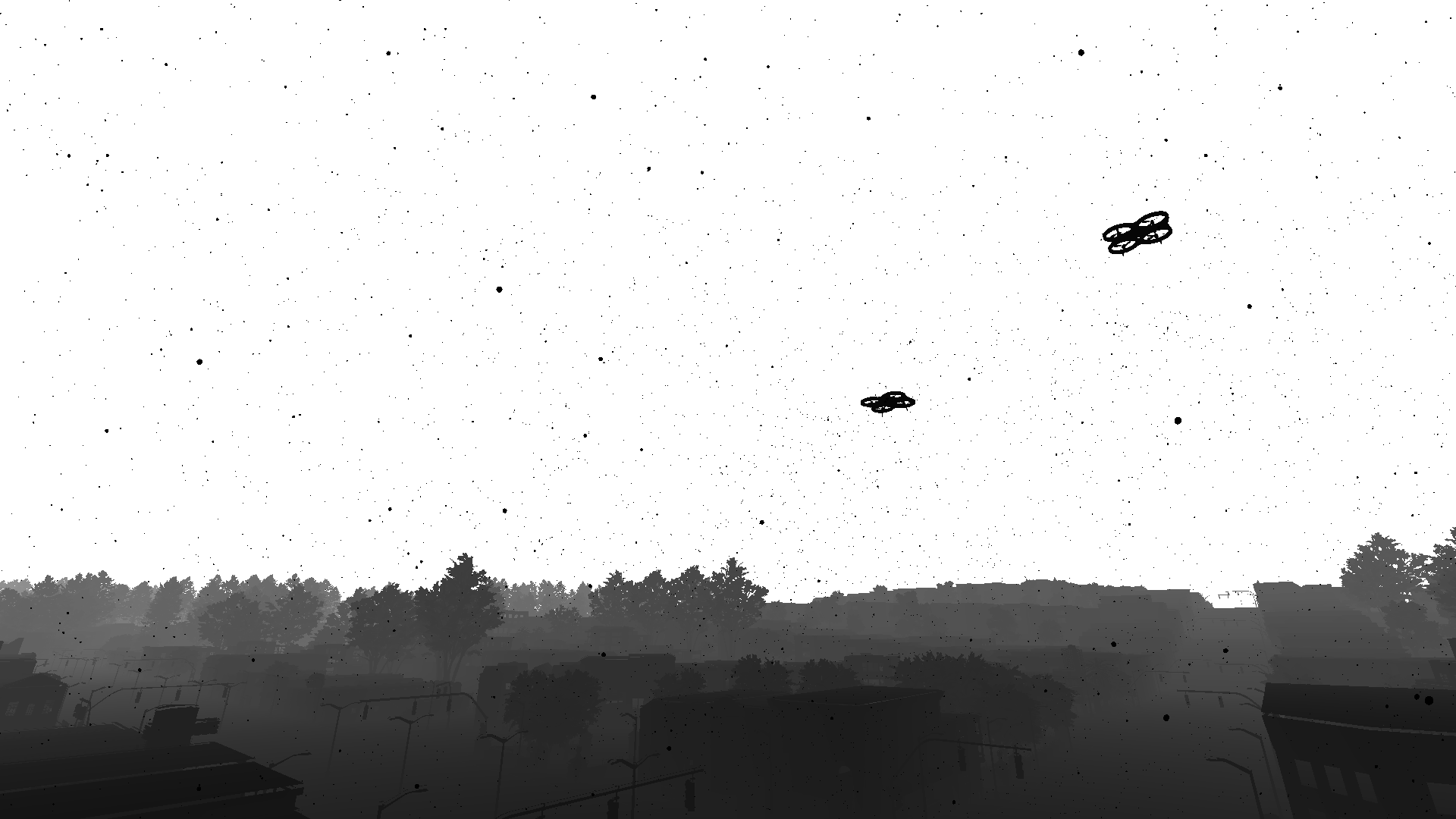} &
\weatherpanel{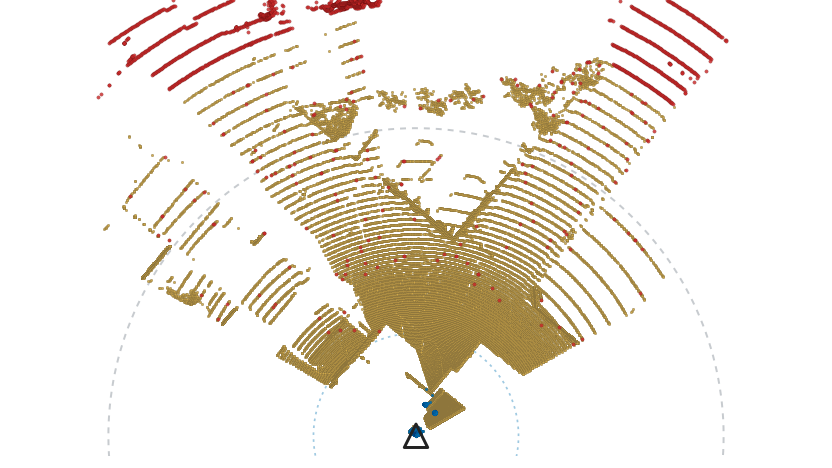} \\
\scriptsize (g) RGB, snowy &
\scriptsize (h) depth, snowy &
\scriptsize (i) LiDAR, snowy \\[0.5mm]
\weatherpanel{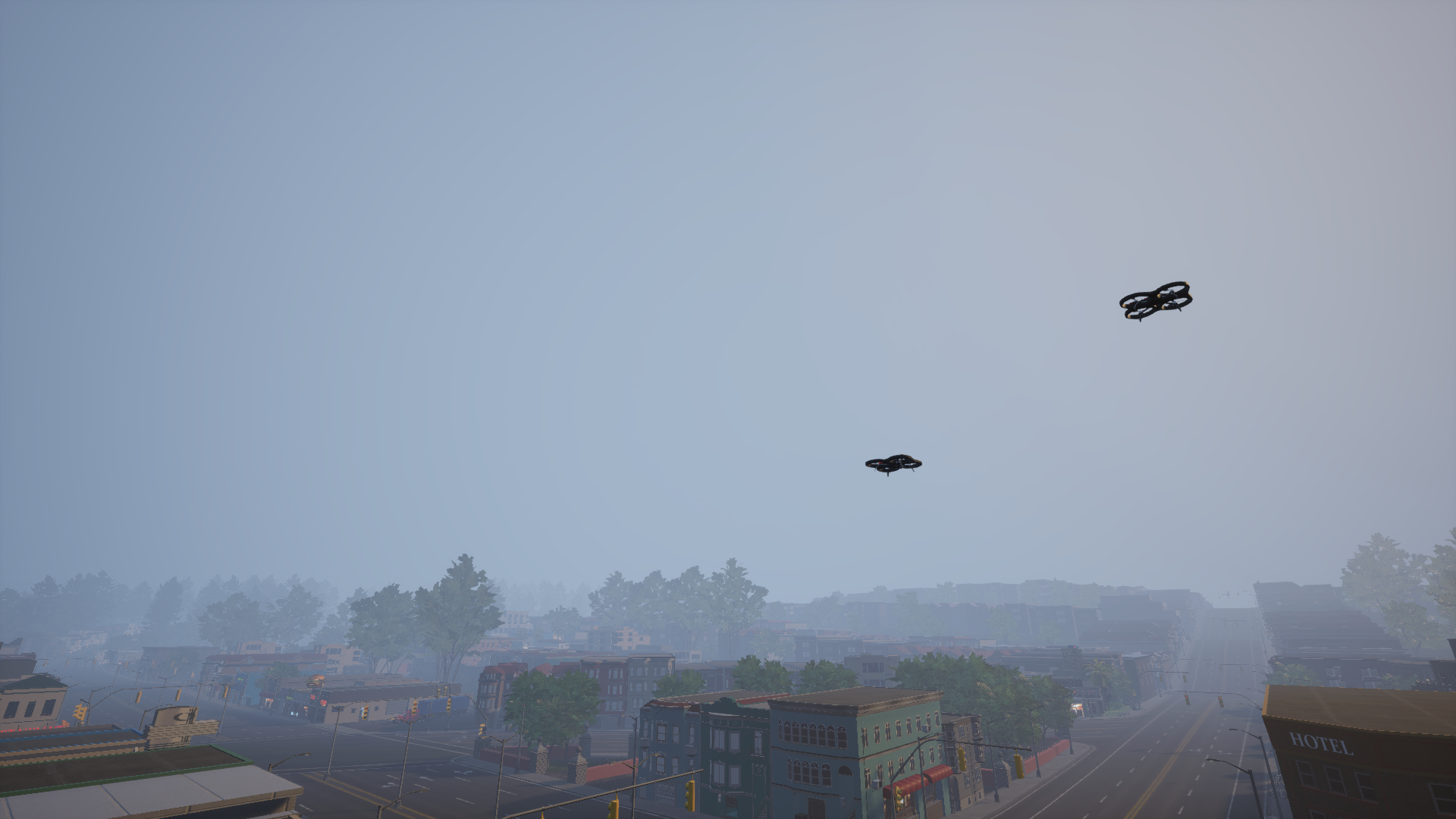} &
\weatherpanel{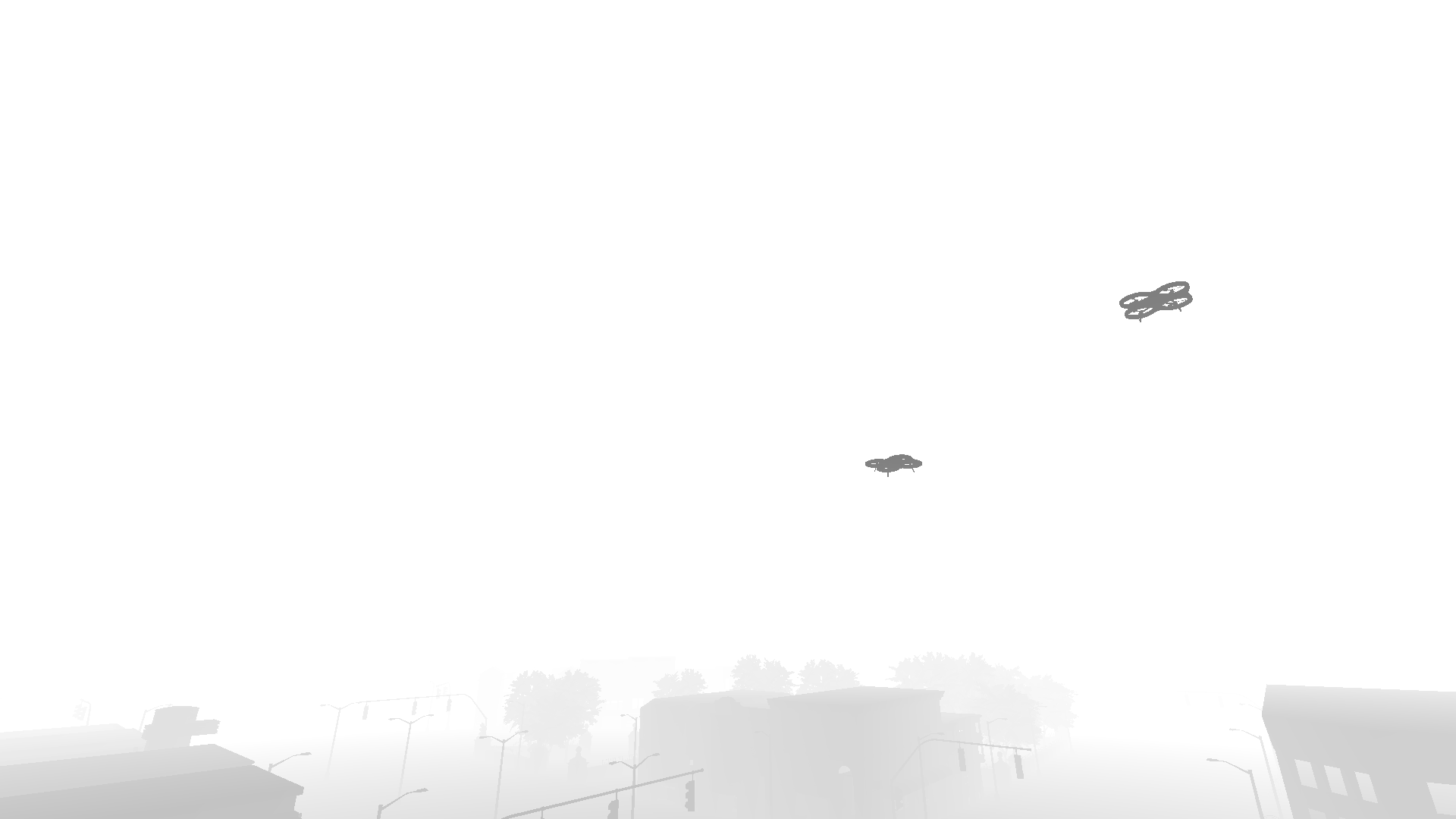} &
\weatherpanel{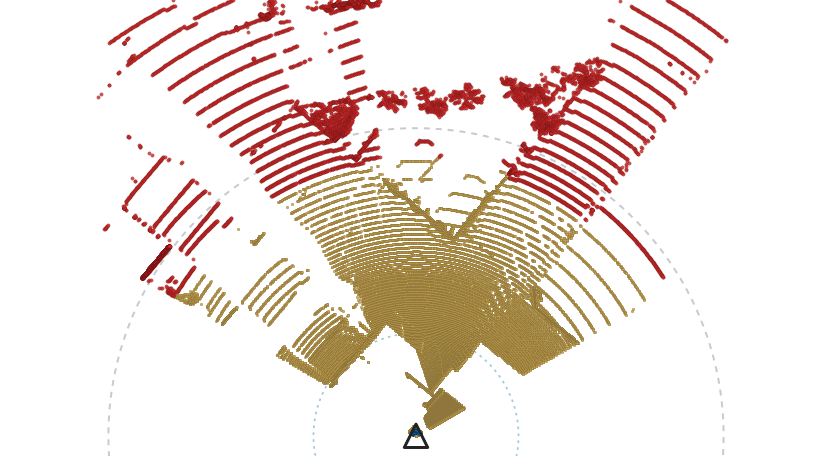} \\
\scriptsize (j) RGB, foggy &
\scriptsize (k) depth, foggy &
\scriptsize (l) LiDAR, foggy 
\end{tabular}}
\caption{Weather visualization examples from the \dataset{} \texttt{Suburb\_1} scene.}
\label{fig:weather_visualization}
\end{figure}

\parahead{Weather Visualization} 
Fig.~\ref{fig:weather_visualization} compares RGB, depth, and LiDAR outputs under matched weather settings. 
The sunny, rainy, snowy, and foggy rows are generated under the same visualization protocol, while the columns compare RGB appearance, depth map, and LiDAR point cloud degradation.
The LiDAR panels keep the original top-down point samples and use red points in the rainy, snowy, and foggy rows to mark sunny returns that are missing in each adverse-weather point cloud, together with blue points for near-range returns caused by raindrops, snowflakes or fog droplets. 
The comparison shows that the weather configuration produces modality-consistent degradation: precipitation and fog change the RGB/depth appearance in the expected direction, while the LiDAR point clouds show both missing sunny-weather points and near-range weather-induced returns rather than unrelated random artifacts.

\begin{figure}[tb!]
\centering
\setlength{\tabcolsep}{3pt}
\renewcommand{\arraystretch}{0.95}
\begin{tabular}{cc}
\includegraphics[width=0.48\textwidth]{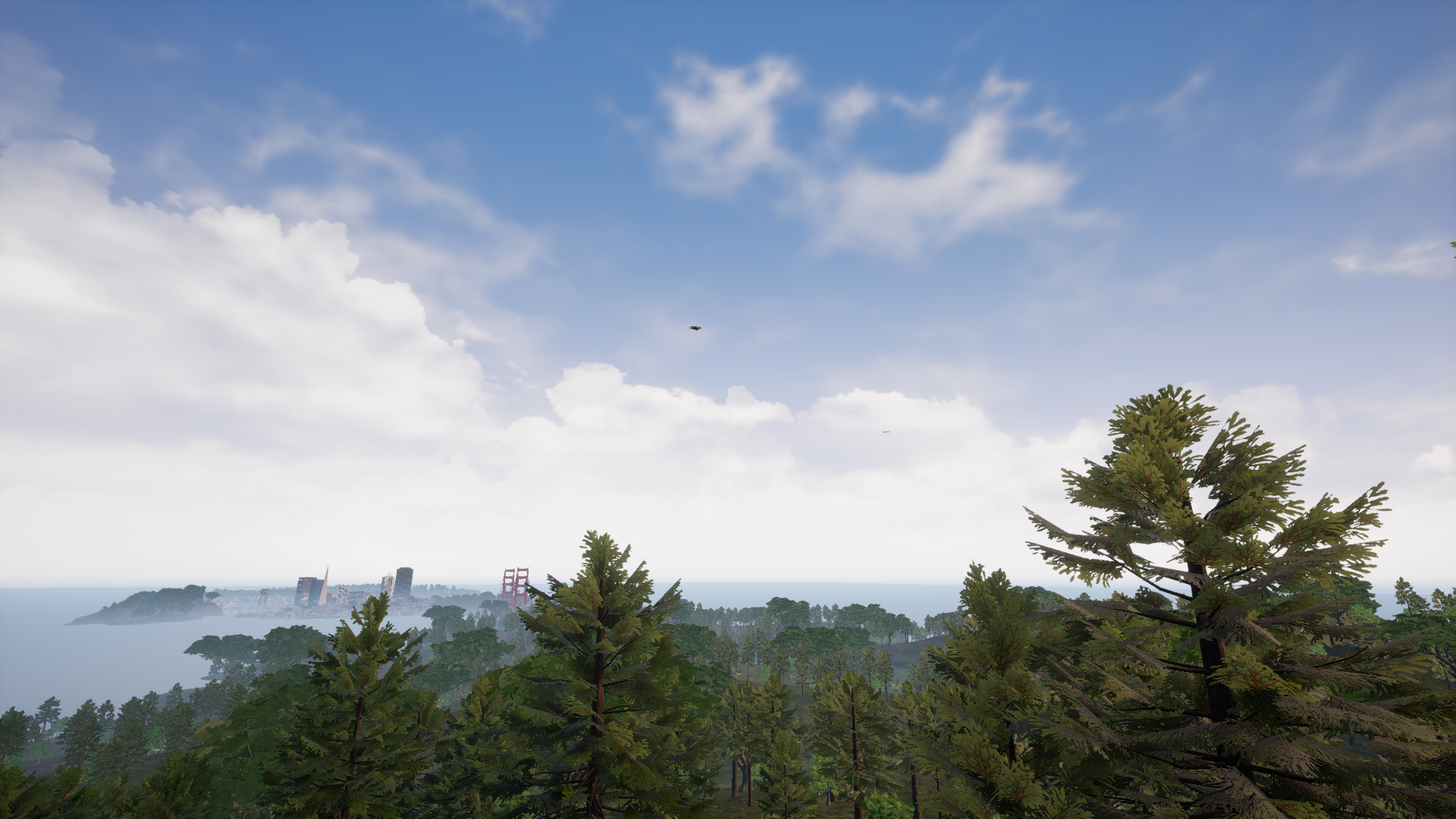} &
\includegraphics[width=0.48\textwidth]{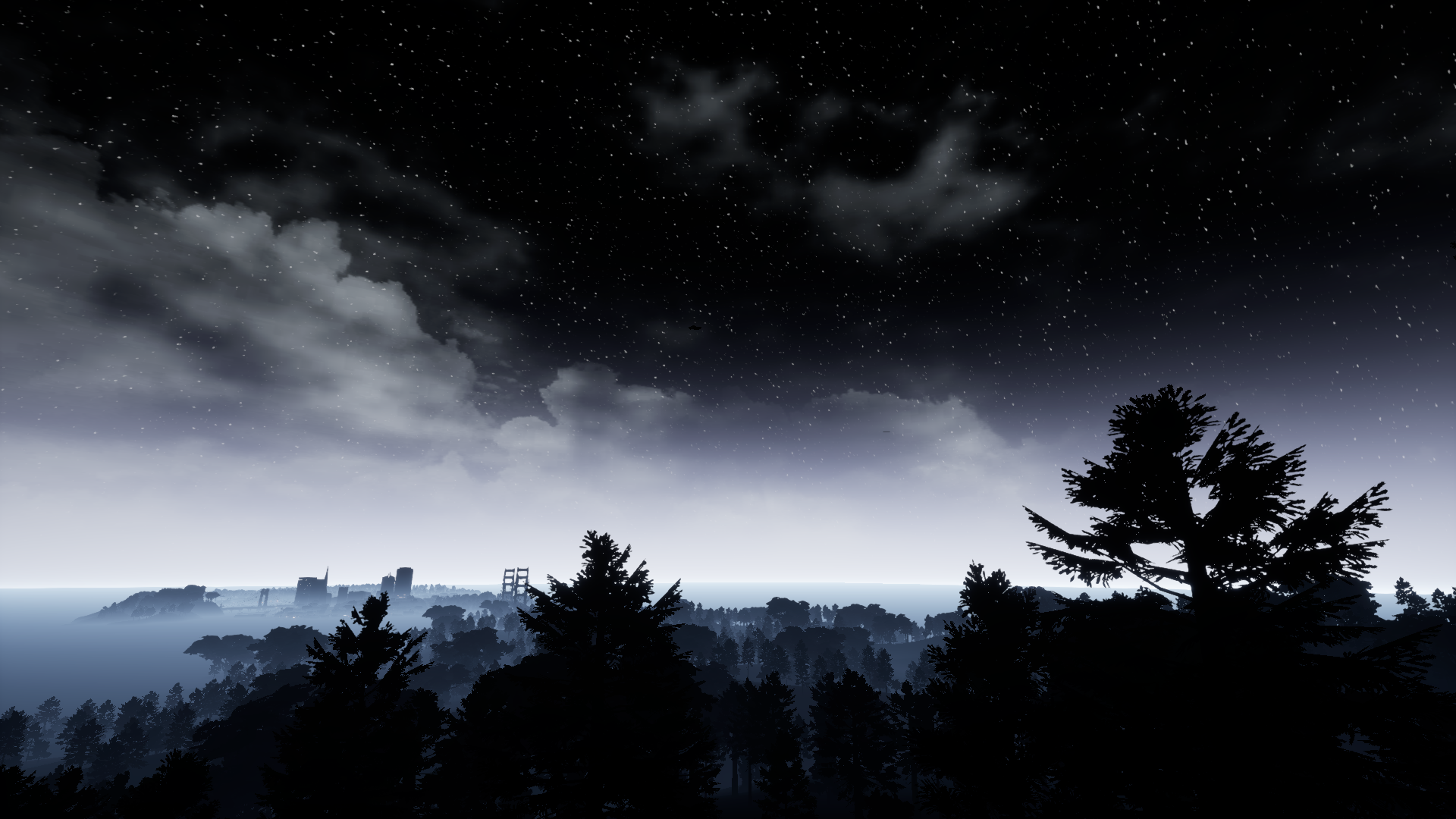} \\
\scriptsize (a) RGB, sunny & \scriptsize (b) RGB, night
\end{tabular}
\caption{Sunny and night visualization example from the LAMBDA \texttt{Mountain\_1} scene.}
\label{fig:illumination_examples}
\end{figure}

Fig.~\ref{fig:illumination_examples} complements this condition-level inspection with a sunny/night pair from the \texttt{Mountain\_1} scene.
This layout helps verify that the weather configuration changes the visible scene in the expected direction, that precipitation and fog effects remain compatible with depth rendering, and that LiDAR degradation is spatially reasonable instead of producing unrelated random artifacts.
These checks complement the frame-level timestamp checks by showing that synchronized sensing streams remain visually and geometrically coherent across weather settings.

\FloatBarrier

\parahead{Synchronized Multimodal Visualization} 
\begin{figure}[tb!]
\centering
{\setlength{\tabcolsep}{2pt}
\renewcommand{\arraystretch}{0.95}
\newcommand{\samplepanelh}[2]{
\parbox[c][#1][c]{0.32\textwidth}{
\centering\includegraphics[width=0.32\textwidth,height=#1,keepaspectratio]{#2}}}
\newcommand{\samplepanelw}[2]{
\parbox[c][#1][c]{0.32\textwidth}{
\centering\includegraphics[width=0.32\textwidth]{#2}}}
\begin{tabular}{ccc}
\samplepanelw{0.21\textwidth}{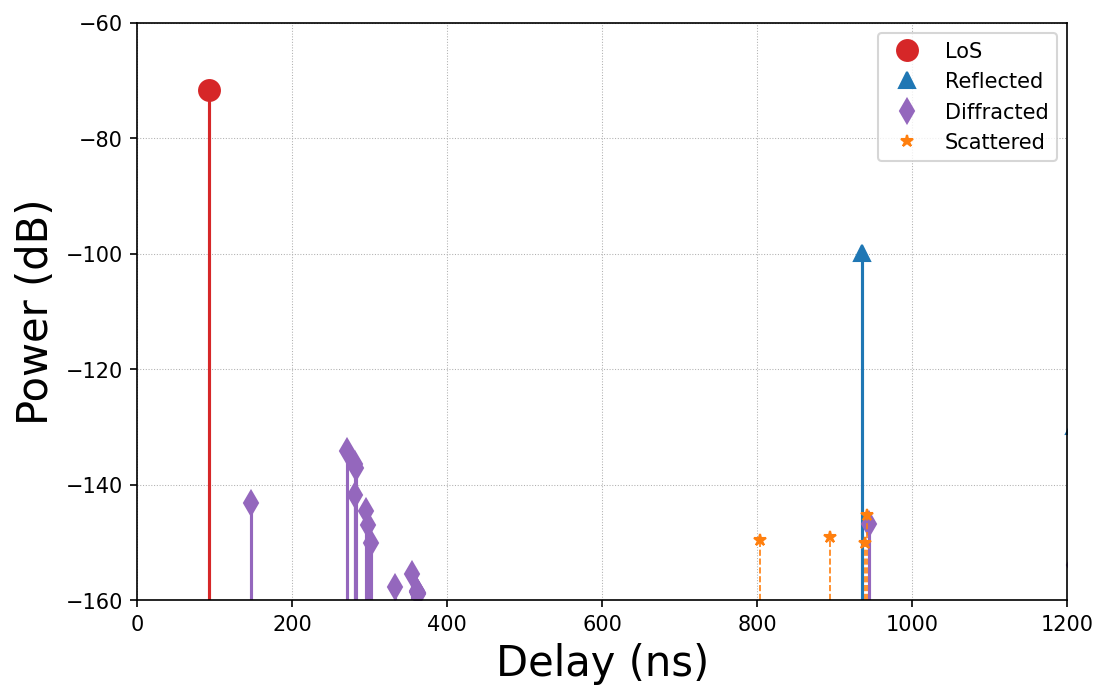} &
\samplepanelw{0.215\textwidth}{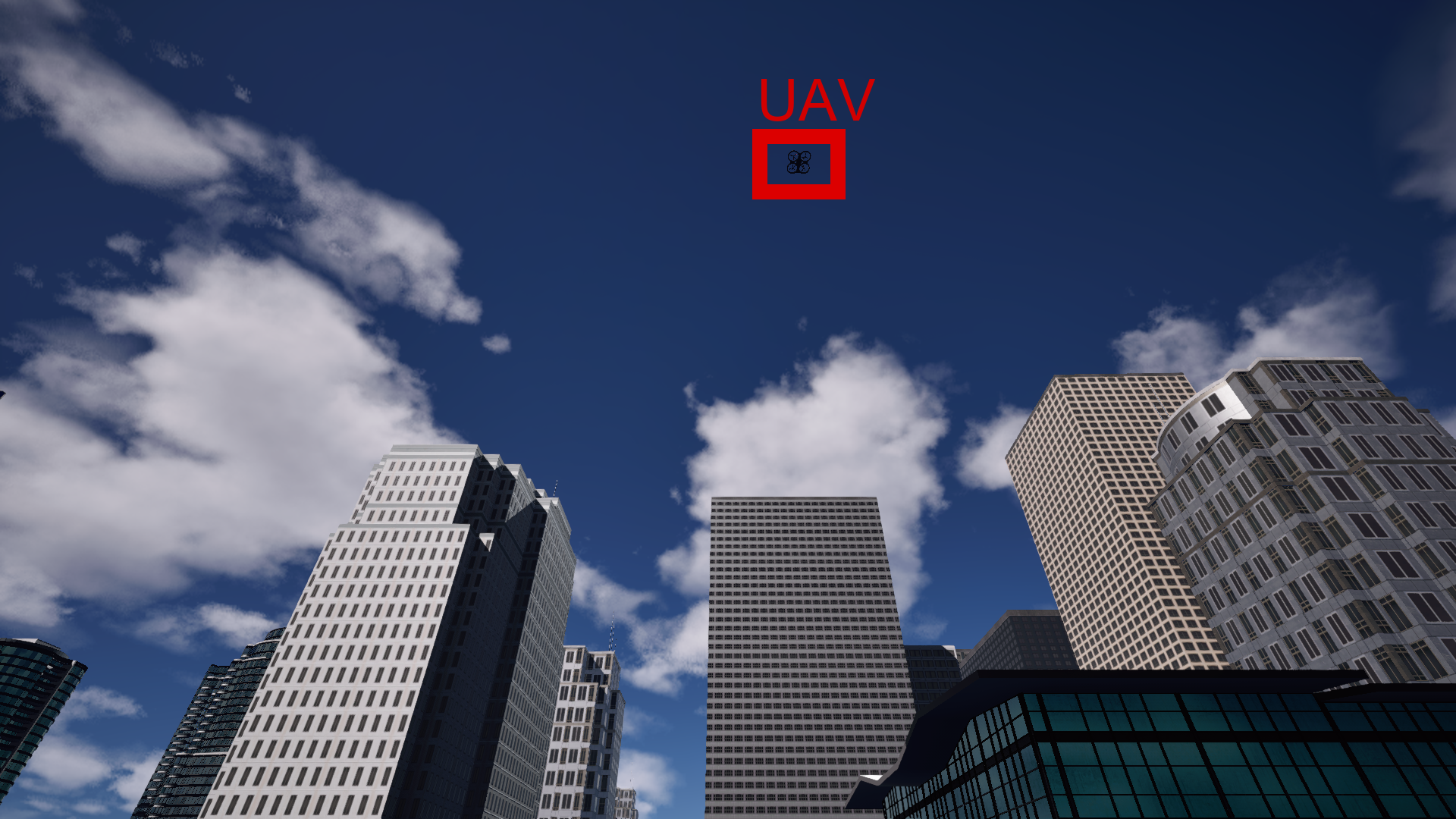} &
\samplepanelw{0.215\textwidth}{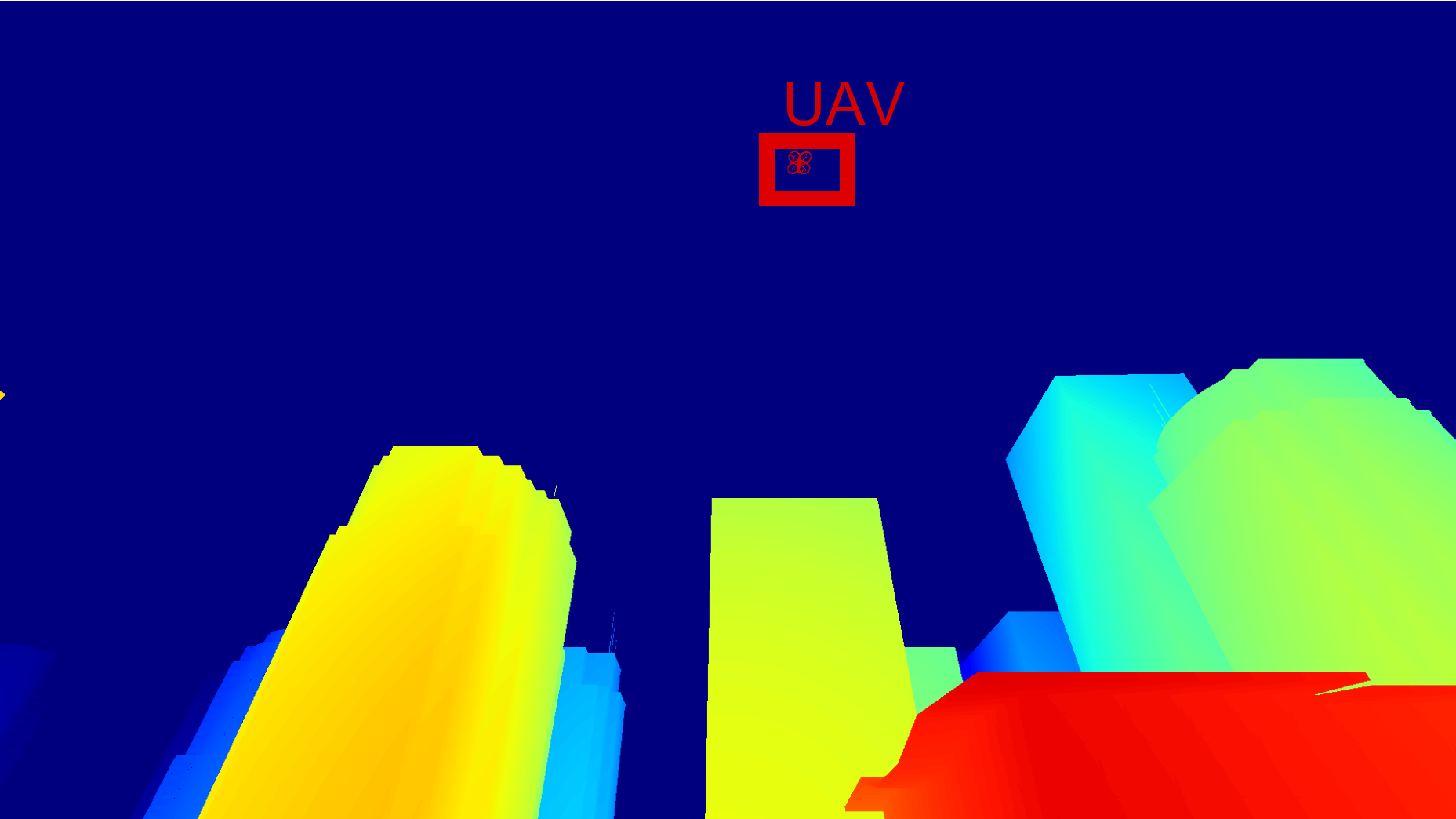} \\
\scriptsize (a) Power-delay profile &
\scriptsize (b) RGB image &
\scriptsize (c) Depth map \\[2mm]
\samplepanelh{0.215\textwidth}{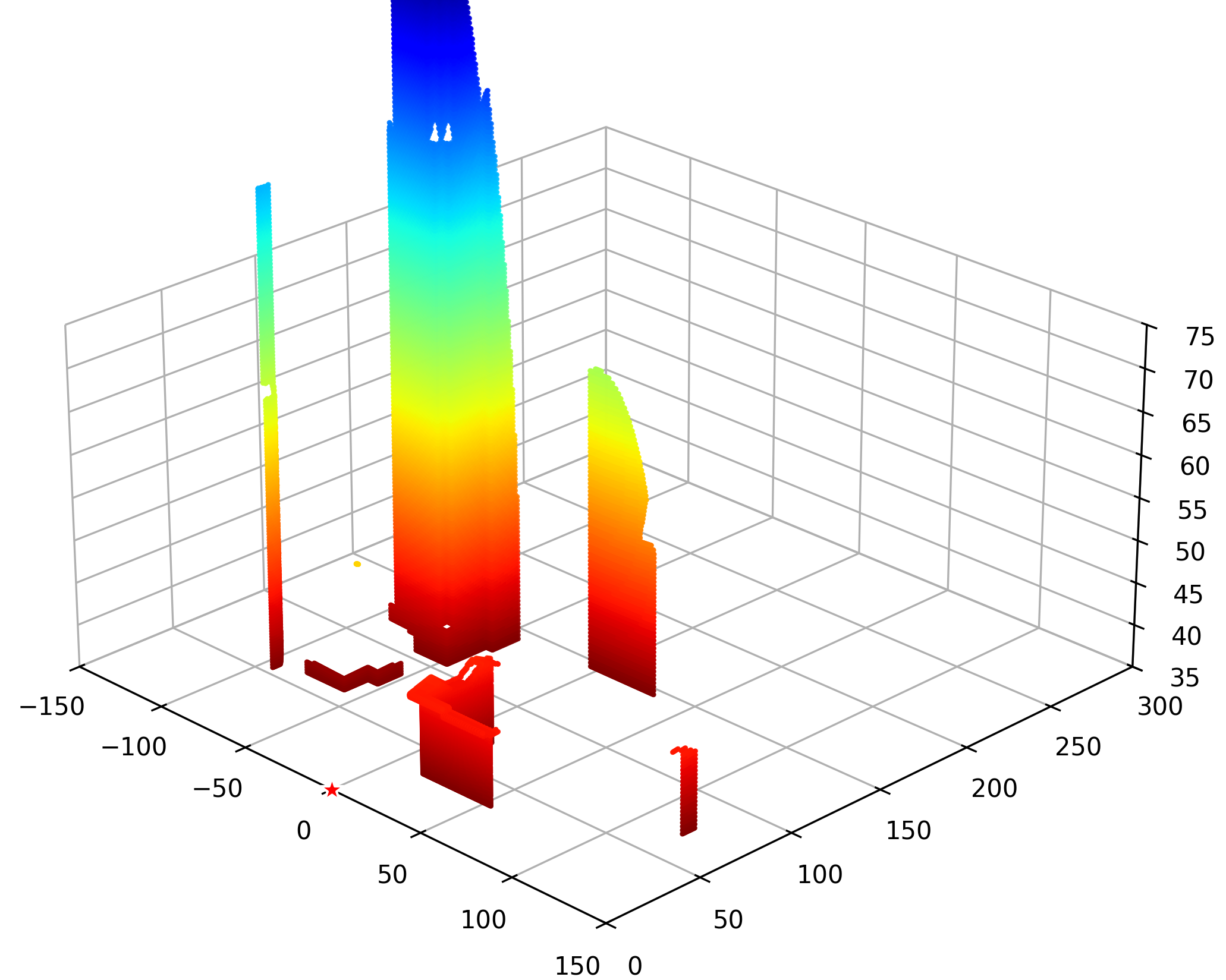} &
\samplepanelh{0.215\textwidth}{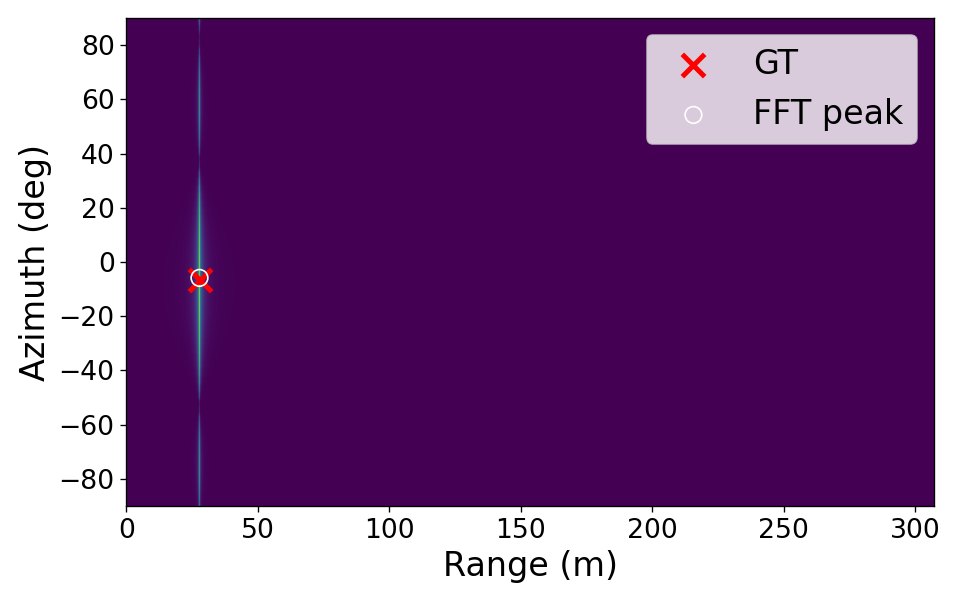} &
\samplepanelh{0.215\textwidth}{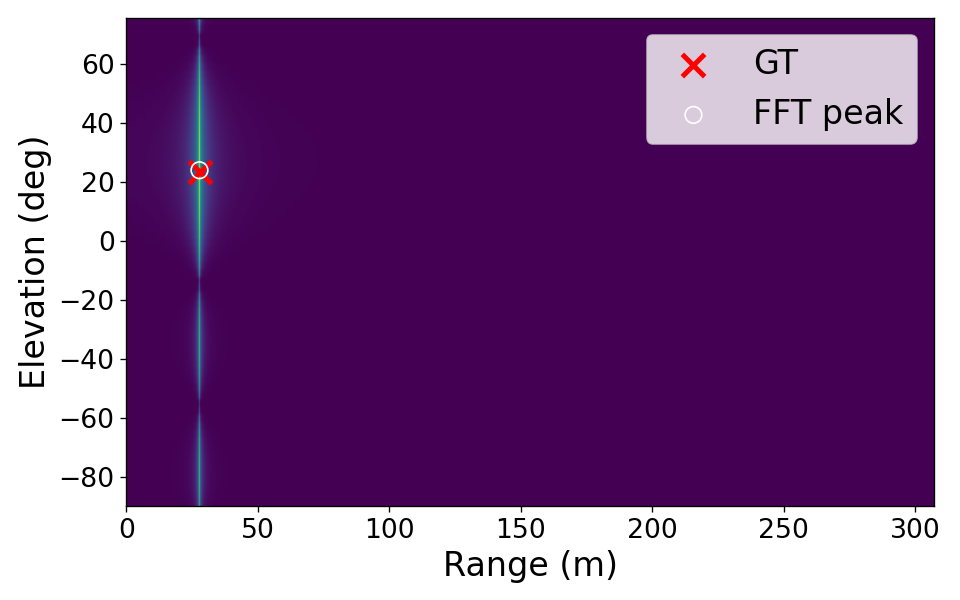} \\
\scriptsize (d) LiDAR point cloud &
\scriptsize (e) Range-azimuth map&
\scriptsize (f) Range-elevation map
\end{tabular}}
\caption{Synchronized multimodal sample from the \dataset \texttt{Block\_1} scene.}
\label{fig:multimodal_sample}
\end{figure}

Fig.~\ref{fig:multimodal_sample} shows a representative synchronized sample loaded from the same frame index in the \dataset{} \texttt{Block\_1} scene.

The synchronized sample is used to visually inspect whether the wireless channel, visual, geometric, and radar-synthesis outputs describe the same low-altitude event.
The power-delay profile is checked together with RGB appearance, depth structure, LiDAR geometry, and radar range-azimuth/elevation angle maps for the same frame index.
This qualitative validation is used to detect frame-index mismatches, stale sensor outputs, obvious field-of-view errors, and inconsistent radar visualizations or CSI products before dataset deposition.

\parahead{Validation via UAV ISAC Use Cases} We further conduct two UAV ISAC experiments to verify that \dataset{} provides learnable and physically consistent supervision signals.
Together, they test whether \dataset{} can be used directly in standard machine learning pipelines.

\begin{figure}[tb!]
\centering
\setlength{\tabcolsep}{0pt}
\renewcommand{\arraystretch}{0.95}
\begin{tabular}{c}
\includegraphics[width=1\textwidth]{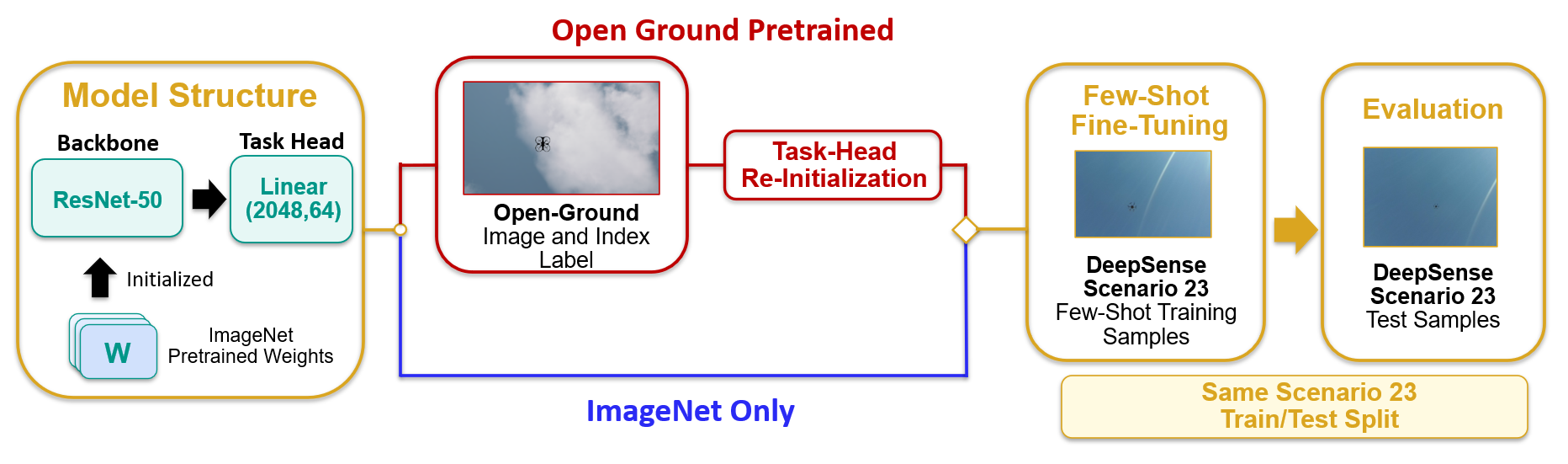} \\
\scriptsize (a) RGB-aided beam prediction workflow. \\[2mm]
\includegraphics[width=0.7\textwidth]{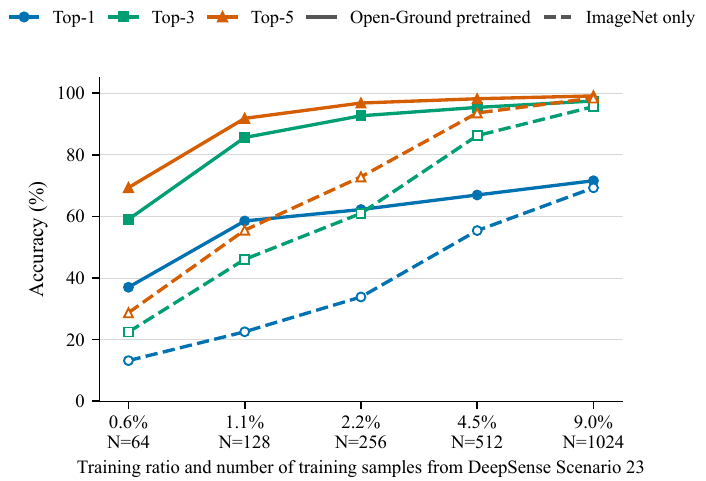}\\
\scriptsize (b) RGB-aided beam prediction results. \\[2mm]

\end{tabular}
\caption{Algorithm workflow and results for Use Case 1: RGB-aided beam prediction.}
\label{fig:isac_use_case_1}
\end{figure}

\taskhead{\textbf{Use Case 1: RGB-aided Beam Prediction}}
The first experiment evaluates RGB-aided beam prediction, whose workflow is shown in Fig.~\ref{fig:isac_use_case_1}(a).
The use case is to infer, from a single RGB image collected at the BS side, which beam in a predefined 64-beam codebook should be selected for the 60 GHz BS--UAV link.
In practical beam management this corresponds to using visual context to speed up or replace exhaustive beam sweeping, because the image provides information about the UAV location and surrounding blockage geometry before the RF beam is selected.

For this experiment, each synchronized RGB--CSI sample is converted into an RGB-aided beam prediction sample.
The 60 GHz CSI is projected onto a 64-beam codebook and the beam with the maximum energy is used as the class label.
The 64-beam codebook is adopted to align the label space with Scenario 23 of the DeepSense 6G real-world dataset~\cite{alsabah2023deepsense}, where the beam prediction annotation is also represented by discrete beam indices.

To keep the validation aligned with a commonly used RGB-aided UAV beam prediction framework, the network architecture follows the protocol used for RGB-aided drone beam prediction in \cite{charan2022dronebeam}, using a ResNet-50 backbone followed by a fully connected classification head. Specifically, the ResNet-50 backbone initialized with ImageNet weights is trained on the \dataset{} \texttt{Open\_Ground} scene.
This source model is trained on 2,639 samples and validated on 293 held-out samples.
For the external transfer check on DeepSense Scenario 23, this backbone pretrained on LAMBDA \texttt{Open\_Ground} is compared with a control model that uses only ImageNet-pretrained backbone weights.
To make the comparison fair, the classification heads of both models are randomly reinitialized before few-shot adaptation to DeepSense Scenario 23, so the results mainly show the effect of \dataset{} \texttt{Open\_Ground} source pretraining on the backbone representation.

Fig.~\ref{fig:isac_use_case_1}(b) illustrates the RGB-aided beam prediction validation results.
It reports the Top-\(k\) accuracy on DeepSense Scenario 23 after few-shot adaptation.
Here, Top-\(k\) accuracy measures whether the ground-truth best beam is included in the top-\(k\) visually inferred candidate beams. 
Solid lines correspond to models initialized with \dataset{} \texttt{Open\_Ground} pretraining, whereas dashed lines correspond to ImageNet-only initialization. 
The models pretrained on \dataset{} \texttt{Open\_Ground} converge effectively, confirming that the synthetic visual-wireless representations provided in the dataset are physically consistent and can serve as valid supervisory signals for downstream tasks.
These results support the intended use of \dataset{} as a low-altitude multimodal pretraining dataset.

\begin{figure}[tb!]
\centering
\setlength{\tabcolsep}{0pt}
\renewcommand{\arraystretch}{0.95}
\begin{tabular}{c}
\includegraphics[width=1\textwidth]{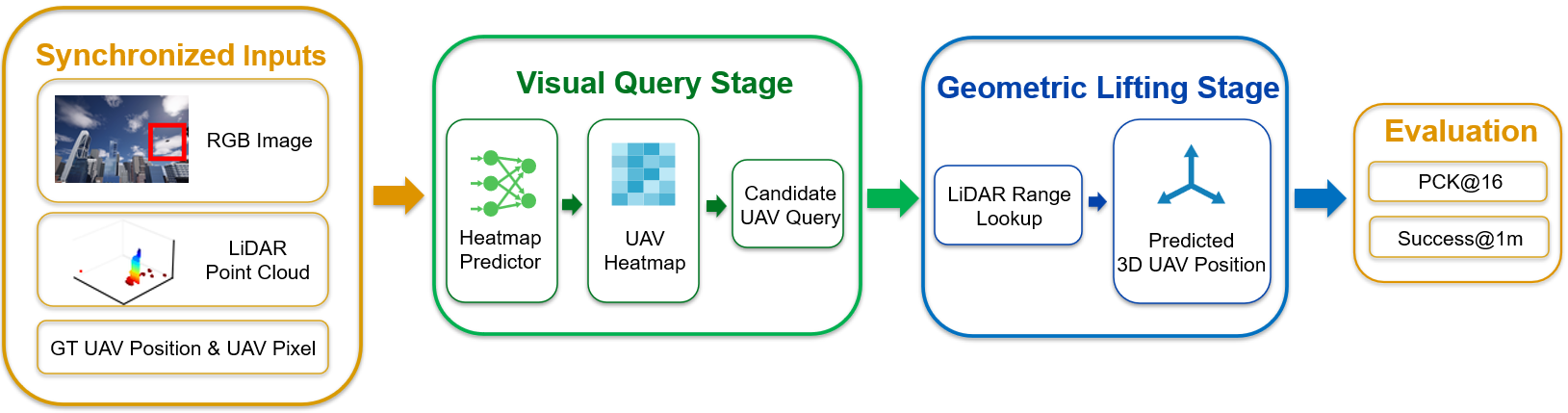} \\
\scriptsize (a) RGB--LiDAR-based UAV localization workflow. \\[2mm]
\includegraphics[width=0.65\textwidth]{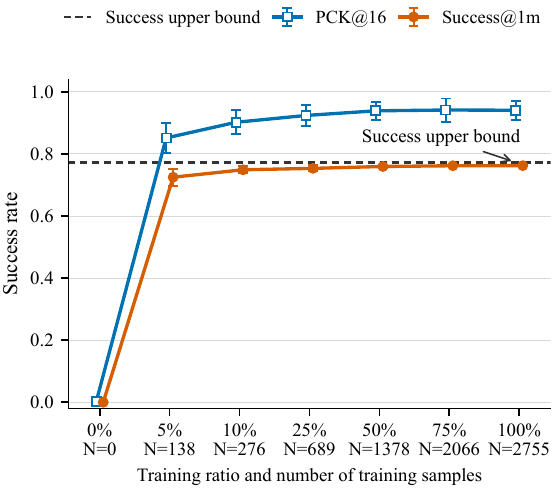}\\
\scriptsize (b) RGB--LiDAR-based UAV localization results. \\[2mm]
\end{tabular}
\caption{Algorithm workflow and results for Use Case 2: RGB--LiDAR-based UAV localization.}
\label{fig:isac_use_case_2}
\end{figure}

\taskhead{\textbf{Use Case 2: RGB--LiDAR-based UAV Localization}}
The second use case evaluates cross-scene 3D UAV localization using synchronized BS-side RGB images, LiDAR point clouds, and ground-truth UAV position and pixel labels, as illustrated in Fig.~\ref{fig:isac_use_case_2}(a). 
The training set is constructed from all available \texttt{Block\_1} samples with UAV heights from 60 to 120 m. 
Testing is conducted on \texttt{Square\_1}; we select the 70--110 m height range for evaluation, with 500 samples at each height and 2,500 test samples in total.

The workflow follows two stages. 
In the visual query stage, a lightweight heatmap predictor takes a \(512\times512\) RGB image as input and generates a \(128\times128\) UAV heatmap. 
The heatmap is supervised by the ground-truth UAV pixel, which is obtained by projecting the frame-level UAV position onto the BS-side camera image using the recorded camera intrinsics and extrinsics. 
During inference, local peaks in the predicted heatmap are extracted as candidate UAV queries. 
In the geometric lifting stage, each candidate query is matched with LiDAR points projected onto the image plane through a LiDAR range lookup. 
The selected 2D query and its associated LiDAR range are then back-projected through the camera calibration and BS pose to obtain the predicted 3D UAV position. 
This pipeline directly tests whether \dataset{} provides consistent RGB labels, LiDAR geometry, camera calibration, and pose alignment across modalities.

For evaluation, we report PCK@16 and Success@1m, as shown in Fig.~\ref{fig:isac_use_case_2}(b). 
PCK@16 measures whether the predicted UAV location falls within a 16-pixel tolerance in the image plane, while Success@1m measures whether the final 3D localization error is within 1 m. 
The dashed line in Fig.~\ref{fig:isac_use_case_2}(b) denotes the Success@1m upper bound obtained by replacing the predicted 2D query with the ground-truth UAV pixel while keeping the same LiDAR range lookup and 3D back-projection procedure. 
Thus, this upper bound reflects the maximum 3D localization success rate allowed by LiDAR coverage and the geometric lifting rule.

The results demonstrate that reliable visual queries are essential for RGB--LiDAR localization. 
Without training, the model achieves only \(0.0020\pm0.0025\) PCK@16 and \(0.00008\pm0.00018\) Success@1m, indicating that LiDAR range lookup alone cannot localize the UAV. 
With only 5\% of the training samples, the model reaches \(0.8512\pm0.0480\) PCK@16 and \(0.7248\pm0.0275\) Success@1m, showing that \dataset{} provides effective visual supervision even in a low-data regime. 
As the training ratio increases from 5\% to 100\%, both metrics improve and then gradually saturate. 
With the full training set, the model achieves \(0.9405\pm0.0301\) PCK@16 and \(0.7621\pm0.0025\) Success@1m, with Success@1m reaching 98.6\% of the upper bound.

Together with the beam prediction experiment, these results show that \dataset{} contains usable RGB--CSI and RGB--LiDAR--position supervision for communication and sensing tasks, while leaving model optimization itself outside the scope of the dataset paper.

\section{Usage Notes}
\label{sec:usage-notes}

\dataset{} can support a broad range of low-altitude UAV sensing and communication studies, including multimodal detection, tracking, localization, channel modeling, beam management, radar-label generation, and cross-modal representation learning. The modalities are spatiotemporally aligned through the shared frame index and the absolute UAV position recorded for each frame. The CSI archives store physically interpretable path-level channel records rather than precomputed OFDM tensors. Users requiring CSI can readily construct it from the stored path-level channel records under their own antenna array size, bandwidth, and subcarrier spacing assumptions.

\section{Data Availability}
\label{sec:data-availability}

The complete \dataset{} dataset is deposited in Science Data Bank under the DOI \url{https://doi.org/10.57760/sciencedb.36052}.
The project website is available at \url{https://www.lambda6g.net/} and provides a download page with subset links.

\section{Code Availability}
\label{sec:code-availability}

A Git repository is publicly available on GitHub at \url{https://github.com/SJTU-WirelessAI-Lab/LAMBDA}.
This repository provides the public \texttt{lambda\_rf} Python package and command-line utilities for the released \dataset{} data, including scripts and examples for CSI inspection, beam label generation, array and subcarrier CSI postprocessing, radar signal synthesis and radar visualization.

\section*{Acknowledgments}
The authors would like to thank Renjie He, Kejia Bian, Weitao Chen, Yuhe Huang, Lingbang Bu, Ziyan Wang, Xiang Li, Xiangwen Gu, and Xinyu Zhang for technical discussions and implementation support for radar RCS modeling, wireless channel modeling, data collection, and validation.

\section*{Author Contributions}
L.Z. conceived the dataset framework, implemented the generation and visualization pipeline, organized the dataset, and drafted the manuscript.
P.R. and C.Z. contributed to scene preparation, data collection, validation, and revision.
J.M. defined the dataset scope, proposed the integrated tool chain for multimodal data generation, and provided technical guidance on data validation.
M.T. defined the strategic scope of the dataset together with J.M., provided technical leadership, supervised the entire workflow, and proposed key methods for dataset validation.
S.S. and Z.C. made suggestions on weather setup, material configuration, and scene construction.
All authors discussed the results and reviewed the manuscript.

\section*{Competing Interests}
The authors declare no competing interests.
\end{document}